\documentclass[%
 reprint,
 superscriptaddress,
 amsmath,amssymb,
 pre,
]{revtex4-2}

\usepackage{graphicx}
\usepackage{dcolumn}
\usepackage{bm}
\usepackage{hyperref}
\usepackage{xcolor}
\usepackage{appendix}
\usepackage{color}

\begin{document}


\title{Cornerstones are the Key Stones: Using Interpretable Machine Learning to Probe the Clogging Process in 2D Granular Hoppers}

\author{Jesse M. Hanlan}
\thanks{J.M.H. and S.D. contributed equally.}
\affiliation{Department of Physics \& Astronomy, University of Pennsylvania, Philadelphia, PA 19104-6396, USA}

\author{Sam Dillavou}%
\thanks{J.M.H. and S.D. contributed equally.}
\affiliation{Department of Physics \& Astronomy, University of Pennsylvania, Philadelphia, PA 19104-6396, USA}

\author{Andrea J. Liu}
\affiliation{Department of Physics \& Astronomy, University of Pennsylvania, Philadelphia, PA 19104-6396, USA}

\author{Douglas J. Durian}
\affiliation{Department of Physics \& Astronomy, University of Pennsylvania, Philadelphia, PA 19104-6396, USA}
\affiliation{Department of Mechanical Engineering and Applied Mechanics, University of Pennsylvania, Philadelphia, PA 19104}

\date{\today}
\begin{abstract}
The sudden arrest of flow by formation of a stable arch over an outlet is a unique and characteristic feature of granular materials. Previous work suggests that grains near the outlet randomly sample configurational flow microstates until a clog-causing flow microstate is reached.  However, factors that lead to clogging remain elusive. Here we experimentally observe over 50,000 clogging events for a tridisperse mixture of quasi-2D circular grains, and utilize a variety of machine learning (ML) methods to search for predictive signatures of clogging microstates. This approach fares just modestly better than chance. Nevertheless, our analysis using linear Support Vector Machines (SVMs) highlights the position of potential arch cornerstones as a key factor in clogging likelihood. We verify this experimentally by varying the position of a fixed (cornerstone) grain, and show that such a grain dictates the size of feasible flow-ending arches, and thus the time and mass of each flow. Positioning this grain correctly can even \textit{increase} the ejected mass by over 50\%. Our findings demonstrate that interpretable ML algorithms like SVMs can uncover meaningful physics even when their predictive power is below the standards of conventional ML practice. 
\end{abstract}

\maketitle

\section{Introduction}
Granular flows occur across natural and designed systems at a variety of length scales.  Whether the constituent grains are pharmaceuticals~\cite{nedderman_flow_1982}, pedestrians ~\cite{helbing_simulating_2000}, electron vortices in superconductors~\cite{olson_reichhardt_vortex_2013} or agricultural grains~\cite{zuriguel_clogging_2014}, the flows are prone to clogging. When the constituent grains pass through an outlet smaller than a few grain sizes, a stabilizing arch structure may spontaneously form, preventing further flow. Clogging has been studied extensively in controlled settings (hoppers)~\cite{to_jamming_2001,alonso-marroquin_beverloo_2020,caitano_characterization_2021,janda_jamming_2008,janda_clogging_2015,hafez_effect_2021,thomas_fraction_2015,koivisto_effect_2017}, varying parameters such as grain shape, friction, and mechanical stiffness, as well as outlet angle and shape~\cite{hafez_effect_2021,pongo_flow_2021,hong_clogging_2017,tao_soft_2021,harth2020intermittent}. Nevertheless, signatures of imminent clog formation remain elusive. 

There is substantial evidence that flow microstates involving $(D/d)^{n}$ relevant grains near the outlet are sampled randomly until one deterministically leads to a clog~\cite{thomas_fraction_2015, koivisto_effect_2017}. Here, $D/d$ is the ratio of the outlet diameter to the grain diameter, and $n$ is the dimensionality of the system, indicating that these grains are contained in an area ($n=2$) or volume ($n=3$) above the outlet, not only in the arch. This model predicts a non-diverging form of average mass ejected per flow event $ \langle M \rangle  \propto  \exp[{(D/d)^{n}}]$, as well as an exponential distribution of ejected masses, both of which match experimental data well~\cite{to_jamming_2005, zuriguel_jamming_2005, Tang09, thomas_fraction_2015, koivisto_effect_2017}. The form of these clog-forming flow microstates remains unknown, but minimal differences between clogging in air and water suggest that they are primarily determined by grain positions, rather than momenta and contact forces~\cite{koivisto_effect_2017}. 

This picture suggests that the \emph{structure} of clogging microstates is important to the clogging process. Machine learning has been successful in identifying a link between local structure and dynamics in disordered systems such as glassy liquids and granular materials~\cite{cubuk_identifying_2015}, and several types of disordered solids~\cite{cubuk_structureproperty_2017, xiao_identifying_2023}. In these works, however, structure was used to predict localized grain-scale rearrangements, which occur frequently throughout the system. In contrast, clogging involves a larger number $\sim (D/d)^{n}$ of grains, and occurs only once per flow event. This makes the problem both less spatially localized and more difficult to adequately sample.

Here we use machine learning tools to predict clogs from a dataset of over 50,000 flow-to-clogging events obtained using an automated hopper. We analyze positional and momentum flow microstates and find that nonlinear deep learning methods or those that include grain momenta perform only marginally better than linear, grain-position-only methods. All methods completely fail to predict clogging until only a short time prior to clogging, supporting the picture of Poissonian sampling of flow microstates. 

Within that short time, the predictive accuracy of our simplest model, a linear Support Vector Machine (SVM) given solely positional information, is 58\%. This is only marginally higher than random guessing (50\%), an unsatisfactory result by prediction and benchmarking standards. Nevertheless, this model identifies the precise location of potential \textit{cornerstones} of an arch as an important predictor of clogging. We confirm that this correlational observation is causal using experiments with a fixed cornerstone grain. This key grain controls the ejected mass by dictating the range of possible flow-ending arches. 

\section{Methods}
\subsection{Data Collection}
\begin{figure}[t!]
\includegraphics[width=8.5cm]{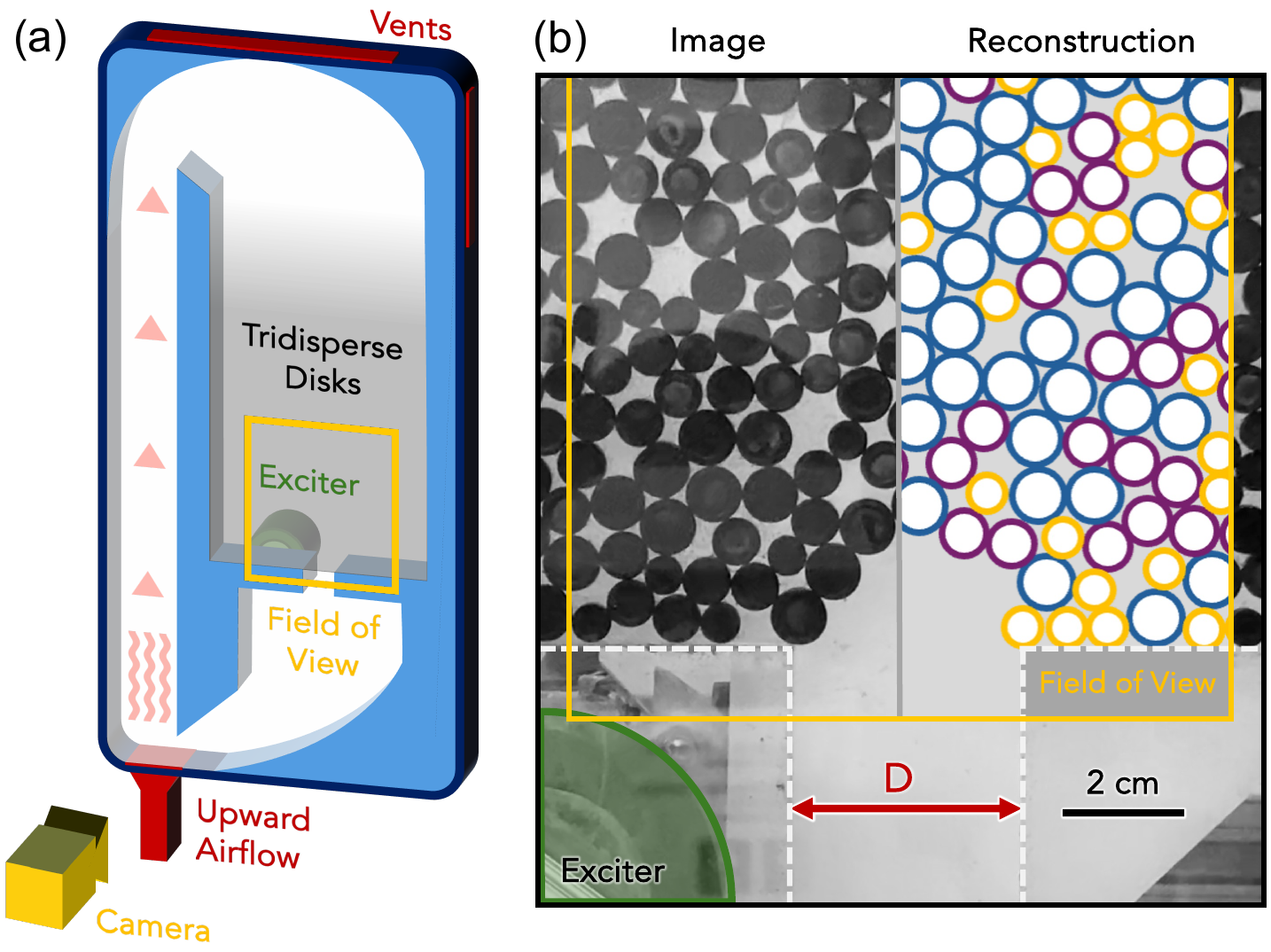}
\caption{\label{fig:Apparatus} (a) Schematic of the automated hopper containing a tridisperse mixture of quasi-2D circular grains (black). Stable arches are broken by an exciter (green) placed behind and below the outlet (raised slightly for visualization, see (b) for exact placement). Grains fall under gravity and are recirculated to the top of the hopper by upward airflow (red) along the left channel. The entire process is recorded by a camera (yellow) at 130 frames per second. (b) Close up of the system near the outlet (left) and schematic of data reconstruction (right). The data recording field of view (yellow) extends beyond the top of this image. $D$ indicates the width of the outlet, which can be varied.}
\end{figure}

We construct an automated quasi-2D hopper (`autohopper'), drawn schematically in Fig.~\ref{fig:Apparatus}a, to directly observe the configurations of grains throughout a flow. The transparent vertical hopper is filled with a single layer of tri-disperse discs of diameters $d_{S} = 6.0$~mm, $d_{M}= 7.4$~mm, $d_{L} = 8.6$~mm, which we will refer to as `grains'. These grains are laser-cut from anti-static Ultra High Molecular Weight Polyethylene (UHMU PE) sheets of mass density $\rho=0.94$~g/cm$^3$ and thickness $h=3.18$~mm. The spacing between front and rear panes of plexiglass is $4.4$~mm so that the grains are free to move but form a monolayer with minimal out of plane displacement. The hopper itself is 22.7~cm wide and 50~cm tall, with fill height of approximately 35~cm.

To begin an experiment, an exciter (green in Fig.~\ref{fig:Apparatus}) situated near the outlet vibrates the hopper, dislodging the arch and initiating flow. The grains then flow freely under gravity until a clog spontaneously forms. The region near the outlet is monitored by a digital camera (yellow) at 130~frames per second. The system is considered stably clogged when no grains have exited the hopper for 5~continuous seconds. For each image taken, custom MATLAB code tracks each grain's size (small, medium, large) and location through time to $\pm\sigma_\mathrm{tracking} = 0.14$~mm precision ($0.016d_L$). This is accomplished prior to starting the next flow, so that tracking data rather than raw video may be written to file to minimize storage requirements. A representation of this process, as well as a stable arch of grains, is shown in Fig.~\ref{fig:Apparatus}b.

Grains that pass through the outlet are directed into a closed loop chute with a blower attached at the base (red in Fig.~\ref{fig:Apparatus}a). An upward airflow recirculates grains to the top of the hopper, removing the need for refilling, and allowing the experiment to continue autonomously without intervention. The air flow is placed sufficiently far and shielded from the outlet such that air currents do not disturb grains in our region of interest, and vents (see Fig.~\ref{fig:Apparatus}a) are placed at the top and sides of the hopper to prevent circulating currents. We perform over 35,000 experiments in this manner for a single outlet size, $D = 3.86d_L$, and at least one thousand experiments each for $D = \{3.61, 3.74, 3.98, 4.15\}d_L$, over 7,000 total. We additionally perform over 13,000 experiments with a fixed particle and outlet size $D = 3.86d_L$ (Fig.~\ref{fig:FixedGrain}).

We confirm a variety of standard granular flow behaviors in Appendix~\ref{HopperPhenom}: the distribution of flow events in exponential (Poissonian), the average event size grows exponentially in $(D/d)^2$, and the average discharge rate follows the 2D Beverloo law. The large quantity of data captured with the autohopper presents a wide range of analysis opportunities. For instance, the dataset contains enough flow events to inform a multiplicative noise model that captures the dynamics of the flow rate and the relative stability of arches~\cite{hathcock_stochastic_2023}. However, for analysis in this work, we restrict our machine learning dataset to a one outlet size, $D = 3.86d_L$, and use the $29,000$ flows that last at least $0.23$~seconds, or $10\%$ of the average flow length. The data for all flows and all outlet sizes is accessible on the Dryad repository~\cite{DataInDryadRepository}. We also provide a Python script to automatically create folders of the expected classes described in the following section~\cite{PythonScript}.

\begin{figure*}[t!]
\includegraphics[width=17cm]{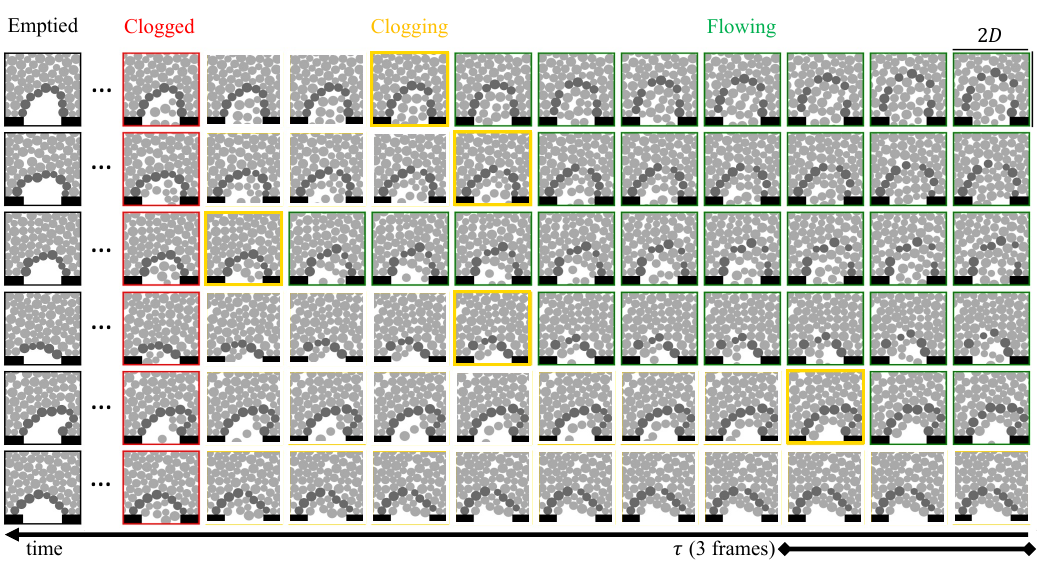}
\caption{\label{fig:StateDef} Still images of six example flow events (rows), labeled by microstate type, which are identified in reverse-chronological order. The in final frame of each experiment, which we label as Emptied (black, left), we identify final arch grains (highlighted). Moving back in time, the clogged frame (red) is the moment in which the arch grains reach their final positions to within tracking precision. The clogging frame (yellow) is the last moment in which the sum of gaps between final arch grains is greater than a small grain diameter $d_S$. All states before the clogging frame are considered flowing (green). The clogging microstate in the bottom row is $9~\tau$ to the right, where $\tau$ is the average time needed for flow microstates to decorrelate. }
\end{figure*}
\subsection{Classifying Microstates}

We approach clogging prediction as a classification problem. To do so, we introduce four classes of flow microstates, and construct a labeled dataset as a ground truth. These classes are \textit{Flowing}, \textit{Clogging} (flow states leading to a clog), \textit{Clogged} (a stable arch has formed), and \textit{Emptied} (all grains have stabilized). By definition these microstates are always experienced in the listed order, though the time spend in each category varies widely. We define these states starting with the emptied state and working backwards. This procedure is described in detail in Appendix~\ref{Labeling}, and briefly, along with six example flows, in Fig.~\ref{fig:StateDef}. The machine learning task is to classify microstates correctly into these four categories. 

\subsection{Machine Learning Analysis}

To be precise, our aim is to use only instantaneous information contained in the microstate (positions, sizes, and momenta of grains) to perform 3 binary classifications to distinguish the flowing state from the clogging, clogged and emptied states, respectively. Thus, our goal is to produce a binary classification function that takes a microstate $\Omega_i$ as input, and produces a single number $\mathcal{C}_i$, which distinguishes between two classes of microstates (\textit{e.g.}  $\mathcal{C}_i<0$ for clogging, clogged or emptied, and $\mathcal{C}_i>0$ for flowing). We compose a function $f$ with many adjustable parameters $\vec \theta$, which we optimize for this purpose using supervised machine learning. Here we assume familiarity with this process, but for an expanded description, see Appendix~\ref{SupervisedLearning}.

Our trainable functions $f$ in this work are primarily linear Support Vector Machines (SVMs)~\cite{burges_tutorial_1998}, but we also train a Convolutional Neural Network (CNN)~\cite{lecun_deep_2015,li_survey_2022} for comparison. We use \textit{hinge loss}~\cite{burges_tutorial_1998,rosasco_are_2004} for the SVMs and \textit{crossentropy loss}~\cite{de_boer_tutorial_2005,lecun_deep_2015,li_survey_2022} for the CNN, with further training details given in Appendix~\ref{SVMFeatureContribution}. We also briefly discuss analysis using Graph Neural Networks (GNNs) in Appendix~\ref{GraphNets}. 

\begin{figure*}[t!]
\includegraphics[width=17.5cm]{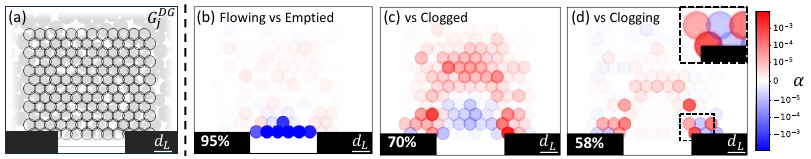}
\caption{\label{fig:SVM_Baseline} Density Grid feature ($G^{DG}_j$) locations (a) and their significance $\alpha_{j}$ for each of the three binary classification tasks: (b) Flowing vs Emptied, (c) Flowing vs Clogged, and (d) Flowing vs Clogging. Features in blue (red) indicate presence of grains in that region is predictive of a flowing (emptied/clogged/clogging) state. The intensity of the color indicates the magnitude of the effect. The areas where individual grain positions matter most are where the gradient of these feature contributions in space is sharpest, as in the the region immediately next to the outlet in (c) and (d).
}
\end{figure*}

In linear SVMs, $f$ takes the form
\begin{equation}
    f(\Omega_{i})_{\vec{\theta}} =\vec{\theta} \cdot \vec G(\Omega_{i})
    \label{SVM_Output}
\end{equation}
where each element of $\vec G (\Omega_i)$ represents a pre-defined feature of microstate $i$. We have investigated several choices of $G$ and present the most informative, $G^{DG}$ (Density Grid), below, with other choices described in Appendix~\ref{PastWork}. In short, each $G^{DG}$ measures the grain density in circular windows arranged on a hexagonal grid, as shown in ~Fig.\ref{fig:SVM_Baseline}a. More precisely, 
\begin{equation}
    G^\mathrm{DG}_{n} = \sum_{\text{grains}} \frac{A_{\text{grain}} \cap A_{n}}{A_n} 
    \quad , \quad G^\mathrm{DG}_0 = 1
\end{equation}
with $A_n = \pi r_{window}^2$, and $\cap \  A_n$ indicates the intersection with the $n$-th circular window. $G^\mathrm{DG}_0 = 1$ gives the system an adjustable offset. We calculate $G_n$ independently for each grain size (small, medium, large), but ultimately find very similar weights assigned for each species. As such, we average significance and feature maps across grain size when displayed in this work. We find varying the spacing and size of circular windows to have negligible effect. For each binary classification, we train our SVM using approximately 20,000 labeled microstates for each class, and report accuracy of classification on a separate test set of approximately 5,000 microstates for each class.

\section{Results}

\begin{table}
\caption{\label{tab:Accuracies} Binary classification accuracy of four machine learning methods distinguishing Clogging, Clogged, and Emptied states from Flowing states. Superscripts DG and BP are for Density Grid and Behler-Parrinello structure functions, respectively.}
\begin{ruledtabular}
\begin{tabular}{cccc}
 Method & Clogging & Clogged & Emptied\\ \hline
 Linear SVM, $G^\mathrm{DG}$ & 58$\%$ & 70$\%$ & 95$\%$ \\
 Linear SVM, $G^\mathrm{BP}$ & 57$\%$ & 68$\%$ & 95$\%$ \\
 Linear SVM, $G^\mathrm{DG}$ (+ Velocity)  & 59$\%$ & 78$\%$ & 99$\%$ \\
 Convolutional Neural Network & 61$\%$ & 84$\%$ & 99$\%$ \\
\end{tabular}
\end{ruledtabular}
\end{table}

\subsection{ML Predictions}
By conventional metrics, our methods perform well separating \textit{Flowing} states from \textit{Emptied} states.  However, separating \textit{Flowing} states from either \textit{Clogged} or \textit{Clogging} states proves difficult, reaching classification accuracies only modestly above chance for the latter. Each of these accuracies are listed in Table~\ref{tab:Accuracies}, along with results using other structure functions ($G^{BP}$), and with added velocity information. We also include results using a far more flexible, nonlinear method, an 830,000~parameter, 35~layer CNN. The details of these additional methods (and several more) are included in Appendix~\ref{PastWork}. Even the most successful method (CNN), the accuracy for the most difficult and important task, distinguishing between \textit{Flowing} vs \textit{Clogging}, is unable to consistently to predict individual clogs, with a test accuracy of only 61\%, which we discuss in detail in Appendix~\ref{Clogginess}. Strikingly, accuracies for this task vary by only 4\% across these methods.  Given this similarity of test accuracy, we focus on the linear SVM that characterizes structure using the density grid. Its simplicity allows us to interpret solutions, and to directly identify structural factors important in clog formation. 

The final weights $\vec \theta$ in the linear SVM have specific spatial importance, that is, they denote where the presence of grains anti-correlates with increased clog likelihood. However to understand our solutions, we must visualize not simply the weights, but the average effect this weight has when applied to the training data. Put another way, the features with greatest variance in their contributions $\sigma_j^2 = \text{var}[\theta_j \times G^{DG}_j(\Omega_i)]_{\text{training set}}$ are those with greatest impact on the decision function, and therefore the most important. We plot \textit{feature significance} $\alpha_{j} = \mathrm{sign(\theta_{j})} \sigma_{j}^{2}$ spatially in Fig.~\ref{fig:SVM_Baseline}b-d. A direct comparison between feature weights $\theta$ and feature significance $\alpha$ can be found in Appendix~\ref{SVMFeatureContribution}.

Despite modest predictive accuracy of the SVM, the feature contributions still give insight into spatial factors of clog formation. First, the prediction of \textit{Emptied} vs \textit{Flowing} states gives an unsurprising feature map in Fig.~\ref{fig:SVM_Baseline}b, where grains (likely falling) in the outlet suggest an \textit{Emptied} state is extremely unlikely. The \textit{Clogged} vs \textit{Flowing} feature significance map in Fig.~\ref{fig:SVM_Baseline}c suggests a relevance of the overall grain density gradient. This may be a means of sensing a slowing flow, occurring at this stage. The fact that velocity information significantly improves the accuracy \textit{only} of the \textit{Clogged} prediction fits nicely with this interpretation (see Table~\ref{tab:Accuracies}).

Notably, when predicting clogging states (Fig.~\ref{fig:SVM_Baseline}d) we see high-valued blue and red regions next to each other at the edges of the outlet. This indicates that moving a cornerstone grain slightly to the right or left might change the prediction drastically. These results suggest that the lateral movements of a single grain in this location may have out-sized importance in clog formation. It is this mechanism that we confirm experimentally in the next section. Further discussion of these significance maps, as well as those using the alternative (Behler-Parrinello ~\cite{behler_generalized_2007}) structure functions are included in Appendix~\ref{PastWork} and Fig.~\ref{fig:SVM_BP}.

\subsection{Experimental Verification of Interpretations}
\begin{figure}[t!]
\includegraphics[width=8.5cm]{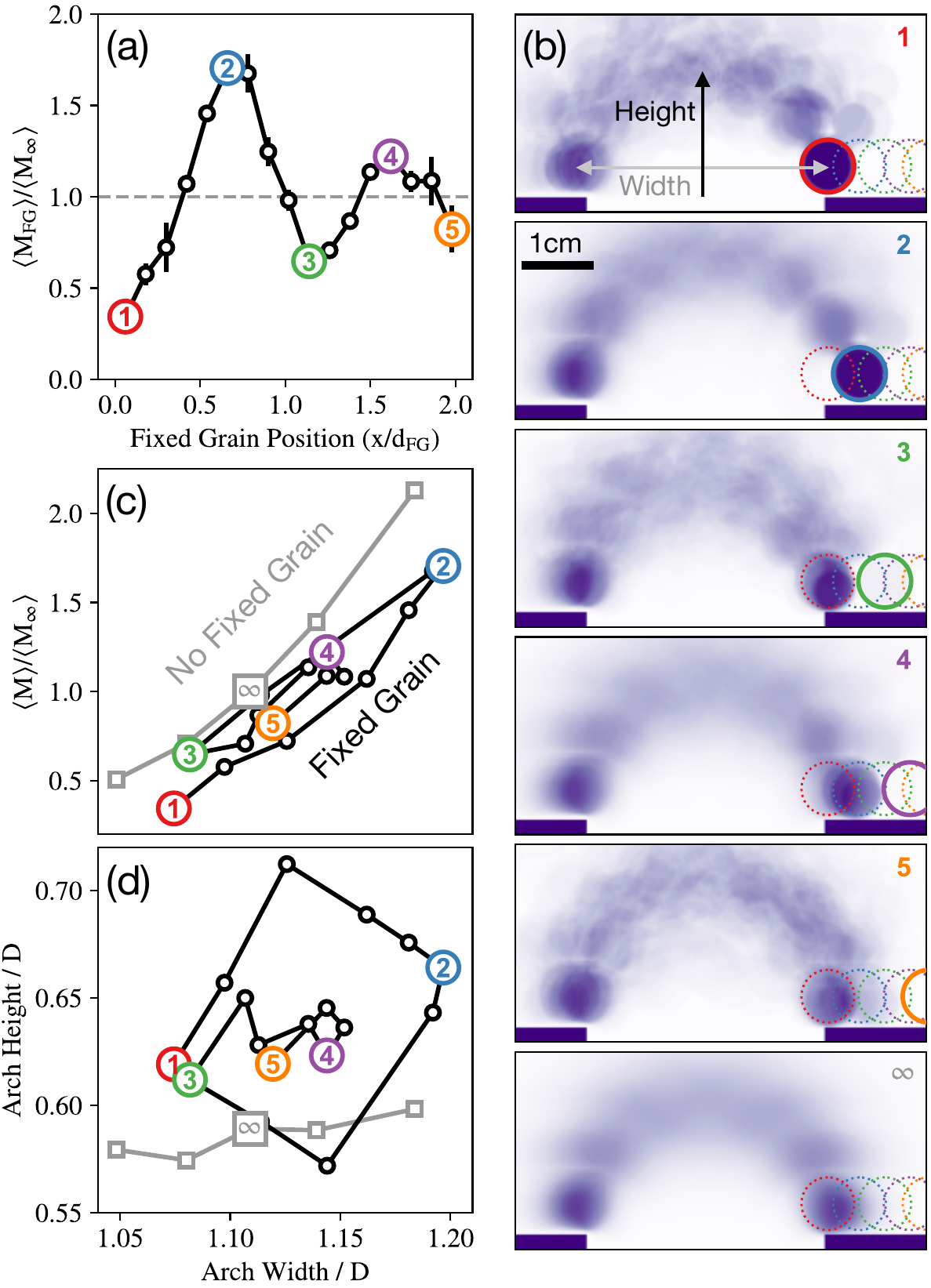}
\caption{\label{fig:FixedGrain} Effect of fixed grain. 
(a) Mean ejected mass $\langle M_{FG} \rangle$ as a function of fixed grain position $x$ relative to the outlet edge. Mass and position are normalized by average ejected mass without a fixed grain $\langle M_\infty \rangle$ and diameter of the fixed grain $d_{FG} = d_{M}$, respectively. Numbered datapoints correspond to maps in (b). 
(b) Averaged final arches for several $x$ values, as well as with no fixed grain (last panel). The relevant fixed grain location is drawn in solid color, and vertical lines are guides for the eye.
(c) Normalized ejected mass vs averaged arch width (horizontal distance between cornerstone centers) normalized by outlet size $D$. 
(d) Arch height vs arch width, both normalized by outlet size $D$. Height is calculated as the vertical distance from the outlet to the highest grain center.}
\end{figure}

Guided by our machine-learned solutions, we experimentally measure the impact of `cornerstone' grain position. We place a fixed grain (magnet) of diameter $\mathrm{d_{FG} = d_{M}}$ on the floor of the hopper near the outlet, as shown by the drawings in Fig.~\ref{fig:FixedGrain}a. This grain is held in place by another magnet on exterior of the hopper. We vary its position $x$ over approximately 7500 experiments, and exclude from analysis any flows where we detect any movement of this grain (fewer than 200).

We find a strong and non-monotonic relationship between the position of the fixed grain $x$ and the resulting average mass flow $\langle M_{FG} \rangle$, as shown in Fig.~\ref{fig:FixedGrain}a. Strikingly, even when the grain does not obscure the outlet ($x>0.5d_{FG}$), its placement may change the average ejected mass by a factor of almost three, including \textit{increasing} the flow above the no fixed-grain case (dashed line in Fig.~\ref{fig:FixedGrain}a). The mechanisms underlying these effects can be understood by visualizing the average final arch grains at several values of $x$, as shown in  Fig.~\ref{fig:FixedGrain}b. 

When obscuring the outlet (small $x$, Fig.~\ref{fig:FixedGrain}b1), the fixed grain serves as the cornerstone of the final arches, which are relatively narrow. As $x$ is increased, the region between the cornerstone and outlet becomes excluded space, unable to stably admit another grain, resulting in wider and wider arches (Fig.~\ref{fig:FixedGrain}b2) and increased ejected mass. At larger distances from the outlet $x > (d_{\text{FG}} + d_{\text{S}})/2 \sim 0.9d_{\text{FG}}$, the fixed grain allows for free-flowing grains to act as a stable cornerstone, resulting in narrower arches (Fig.~\ref{fig:FixedGrain}b3) and reduced ejected mass once again. As $x$ increases further, the fixed grain continues to indirectly dictate cornerstone position, even when it is multiple diameters away from the outlet (Fig.~\ref{fig:FixedGrain}b4 and 5). At this stage, the effect of $x$ is reduced, which we attribute to the random availability of differently-sized cornerstones. Overall, we find a clear correlation between average arch width $A_x$, and the average ejected mass, as shown in Fig.~\ref{fig:FixedGrain}c. This observation dovetails nicely with the Thomas and Durian model~\cite{thomas_fraction_2015}, as wider arches require a larger area of grains to cooperate. As a result, there is a smaller likelihood of clogging per sampling time. We find that arches formed in the presence of a fixed grain are slightly wider and significantly taller than those generated without one, as shown in Fig.~\ref{fig:FixedGrain}d, perhaps a result of the additional stability of the fixed grain.

\section{Discussion}
We have constructed an automated quasi-2D hopper, and performed and analyzed tens of thousands of clogging experiments. By defining four classes of states (flowing, clogging, clogged, emptied), we cast clogging prediction as a machine learning (ML) classification problem. We found that even sophisticated nonlinear methods like Convolutional Neural Networks (CNNs), do not produce reliable identification of clogging vs flowing states. By usual prediction or benchmarking standards, the test accuracy (61\% for CNNs) is too close to random guessing (50\%) for the prediction to be useful. Classification accuracy, however, is not the same as physical insight. 

We have shown that even with low test accuracy we have uncovered new physics using ML. In particular, by inspecting the features of greatest significance in a simpler method, a linear Support Vector Machine (SVM), we were able to identify the region immediately adjacent to the outlet as potentially critical to the onset of clog formation. Guided by this ML analysis, we performed a series of experiments with fixed grains in this key position, and showed that the position of the `cornerstone' grain has a large effect on ejected mass, potentially increasing it by over 50\%. Finally, we showed that this relationship stems from the cornerstone grain's ability to dictate the size of final arches, and thus the clogging likelihood. These results have implications for practical hopper design, and provide an object lesson for utilizing machine learning for scientific exploration.

Our findings suggest a rich set of open questions about this and other granular-flow systems. Our system (and others like it) encounters meta-stable arches frequently, only to spontaneously resume flow~\cite{hathcock_stochastic_2023}. Might portions of the outlet region be continually finding rigid substructures, only to have them fall apart due to lack of cooperation? In a larger view, what is the relative importance of microstate sampling (finding an arch) vs arch stability? Further work with fixed grains might prove useful here, by limiting the arch structures available. 

In this work we have attempted a wide variety ML methods, including many variations of SVMs, high-dimensional linear regression, several CNNs, Graph Neural Networks (GNNs). We also included velocity information, modified the scale of binning of features, and more. The methods not included in the main text are described in detail in Appendix~\ref{PastWork}. None of these attempts yielded appreciably better \textit{Flowing} vs \textit{Clogging} classification accuracy than our linear SVM. Of course, our numerous attempts do not prove there is no better solution, and we encourage other researchers to try their hand in improving upon our benchmarks. To facilitate such a competition we make our data available at~\cite{DataInDryadRepository}. Additionally, we have detailed a variety of alternative analyses on this data and potential pitfalls in Appendix~\ref{FutureWork}. One notable pitfall is the imposition of too much coarse-graining, including prematurely enforcing symmetries, even those imposed by the boundary conditions (such as left/right symmetry). In optimization problems it is often helpful to have additional degrees of freedom to \textit{find} the solution, even if they are ultimately not required~\cite{schaeffer_double_2023}. 

For any practical purposes, our current results fail to accurately predict imminent clogs. However, our success in leveraging these `bad' solutions to uncover the importance of cornerstones highlights the usefulness of interpretable ML methods. Because our ML analysis (Fig.~\ref{fig:SVM_Baseline}) is correlational, however, it alone provides an insufficient basis for any \textit{causal} claims. Solution weights may not have obvious meaning. This is a thorny aspect of ML or any optimization process, as finding good solutions often requires over-parameterization~\cite{schaeffer_double_2023}, which by definition permits meaningless variation in the solution weights $\vec \theta$. As a result, ML analyses are typically restricted to claiming that predictive information is present in the data. This type of claim is not without its scientific uses~\cite{dillavou_quality_2022}, however it does not provide mechanistic understanding. 

In contrast, we have employed an `unsuccessful' ML analysis as a compass needle to guide experiments. The result was causal insight into a rare, nonlinear, collective event (clog formation) that depends on a huge number of degrees of freedom and is influenced by poorly understood processes like frictional aging~\cite{dillavou_quality_2022}. ML methods do not replace causal scientific reasoning and experimental verification. However, our results do suggest that the ability of ML methods to find high-dimensional correlations in messy data can be harnessed, even when they fail to make accurate predictions, to guide experiments studying a broad range of complex phenomena across many fields.

\begin{acknowledgments}
This work was partially supported by NSF grants DMR-1619625, MRSEC/DMR-1720530 and MRSEC/DMR-2309043. SD acknowledges support from the University of Pennsylvania School of Arts and Sciences' Data Driven Discovery Initiative. AJL and DJD thank the Center for Computational Biology at the Flatiron Institute, a division of the Simons Foundation, as well as the Isaac Newton Institute for Mathematical Sciences under the program ``New Statistical Physics in Living Matter" (EPSRC grant EP/R014601/1), for support and hospitality while a portion of this research was carried out. 

\end{acknowledgments}


\appendix

\section{Hopper Phenomenology} \label{HopperPhenom}

\begin{figure}[t!]
    \includegraphics[width=8.6cm]{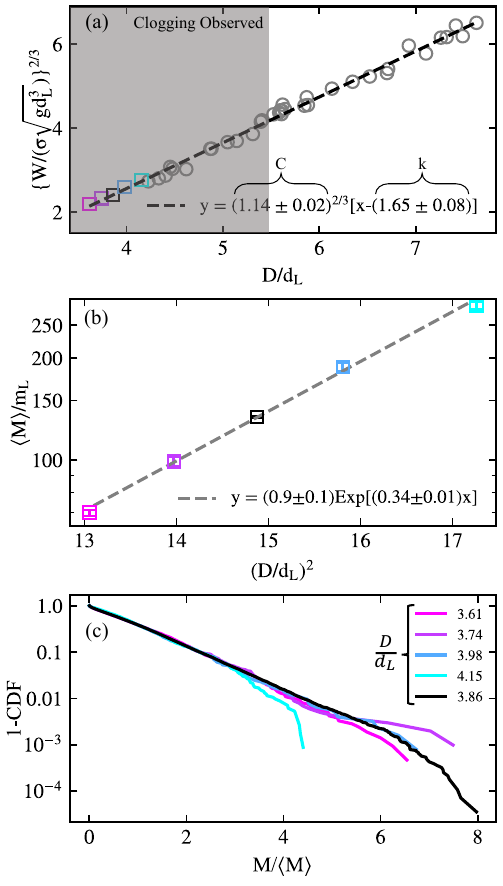}
    \caption{Characteristic behaviors for hopper flow. (a) Flow rate follows the Beverloo equation. Linearized flow rate prediction from the Beverloo equation vs outlet size. Colored squares are averages of $>1000$ flows each taken by our automated hopper, and correspond to data in (b) and (c). Gray circles are each single experiments taken by hand. The shaded region is the approximate regime where we observe clogging. (b) Average ejected mass prior to clog formation versus outlet size squared. We plot the expected exponential relationship with a best fit line. (c) Cumulative Distribution Function (CDF) of the ejected mass, from initiation of flow to clog formation, versus outlet size. In each case the data matches the expected exponential distribution.}
    \label{fig:StandardMeasures}
\end{figure}

Using our automated data, we confirm standard hopper behavior, as shown in Fig.~\ref{fig:StandardMeasures}. For each flow we calculate the ejected mass, $M$ as the total mass of grains that pass a semi-circular boundary spanning and centered on the outlet prior to clog formation. This boundary choice permits ignorance of fast-moving grains in the center of the outlet, which are the most difficult to track.

For each outlet size $D$ we calculate the average flow rate, and find they collapse to the expectation given by the Beverloo equation:
\begin{equation*}
    W = C \sigma \sqrt{g d_{\mathrm{L}}^{3}} (D/d_{\mathrm{L}}-k)^{3/2}
\end{equation*}
as shown in Fig.~\ref{fig:StandardMeasures}a, where $C$ and $k$ are dimensionless fit parameters of order unity, $\sigma=\rho h$ is mass/area of the grains, and $g=9.80$~cm/s$^2$ is the acceleration due to gravity. We include measurements of significantly larger outlet sizes, including those that do not clog, in this plot as gray circles.

We find also that the average ejected mass scales as an exponential of the outlet size raised to the dimensionality of the system $\langle M \rangle \propto \exp[(D/d_L)^2]$, as shown in in Fig.~\ref{fig:StandardMeasures}b for five outlet sizes with $>1000$ flow experiments each. Finally we demonstrate the expected exponential distribution of discharged masses for all outlet sizes, as shown in Fig.~\ref{fig:StandardMeasures}c. Thus our autohopper displays the three hallmark features of a system with granular clogging.

For simplicity of machine learning analysis, we limited our analysis to a single outlet size, $D = 3.86d_L$, shown in black in Fig.~\ref{fig:StandardMeasures}. This outlet has an average flow duration of $\langle t_{\mathrm{flow}} \rangle = 2.26$~seconds and average ejected mass of $\langle M \rangle  = 134m_L$. To avoid transient effects at the initiation of flow, we exclude flows shorter than 10$\%$ of the average flow time, $t_{\mathrm{flow}} < 0.23$~seconds. Thus our machine learning data set uses approximately 25,000 flow events for this single outlet size.

\section{Supervised Machine Learning}\label{SupervisedLearning}
We train a flexible function $f$ with adjustable parameters $\vec \theta$ to distinguish between two classes of microstate by producing a scalar output $\mathcal{C}$. For example, distinguishing \textit{Flowing} vs \textit{Clogging} states would mean
\begin{equation}
  f_{\vec \theta}(\Omega_i) = \mathcal{C}_i
    \begin{cases}
      <0 & \text{if $\Omega_i$ is \textit{Flowing}}\\
      >0 & \text{if $\Omega_i$ is \textit{Clogging}}\\
    \end{cases}     
    \label{funcgoal}
\end{equation}
Where $\Omega_i$ is flow microstate $i$, as specified for example by the positions of grains near the outlet. To find values of $\vec \theta$ that best classify a microstate, we define a a loss function $\mathcal{L}(\mathcal{C}_i,L_i)$ that quantitatively evaluates the performance of $f_{\vec \theta}$ using its output $\mathcal{C}_i$ (which we refer to as \textit{Clogginess}) and our labels $L_i$. The essential feature for a cost function is that it monotonically decreases as our function's output improves. For the example in Eq \ref{funcgoal}, this would equate to 
\begin{equation}
  \frac{d}{d\mathcal{C}_i} \mathcal{L}(\mathcal{C}_i,L_i) 
    \begin{cases}
      \leq 0 & \text{if $\Omega_i$ is \textit{Flowing}}\\
      \geq 0 & \text{if $\Omega_i$ is \textit{Clogging}}\\
    \end{cases}     
\end{equation}
Simply put, if we lower the cost function summed across our entire training (labeled) dataset, we improve the quality of our classification function. Therefore, our aim in training $f$ is to find the parameters $\vec \theta^*$ that minimize our cost function summed over the training dataset of labeled microstates $[\Omega_i, L_i]$:
\begin{equation}
  \vec \theta^* = \text{arg min}_{\vec \theta} \sum_i \left( \mathcal{L}(f(\Omega_i)_{\vec \theta} ,L_i) \right )    
\end{equation}
This is the \textit{supervised learning} paradigm, and may be accomplished using many optimization methods, for example \textit{stochastic gradient descent}\cite{geron_hands-machine_2019, bishop_pattern_2006}, or even explicit minimization for simple forms of $f$. Ultimately, we evaluate and report the performance of our trained functions by the fraction of microstates they label correctly in a reserved \textit{test dataset} not used during training. In this work we utilize two forms of $f$, linear kernel Support Vector Machines (linear SVMs)~\cite{burges_tutorial_1998} and (nonlinear) Convolutional Neural Networks (CNNs)~\cite{lecun_deep_2015,li_survey_2022}. We use two loss functions $\mathcal{L}$, \textit{hinge loss}~\cite{burges_tutorial_1998,rosasco_are_2004} for the SVMs and \textit{crossentropy loss}~\cite{de_boer_tutorial_2005,lecun_deep_2015,li_survey_2022} for the CNNs. Both loss functions are standard choices in machine learning classification problems. 

\section{Labeling and Cleaning Data}\label{Labeling}

We restate the labeling process below, with added details. The emptied state is the final recorded frame of the flow, well after all grains that will leave the hopper have left. From the emptied state, we identify the grains that compose the final arch by tracing the shortest path from left outlet edge to right outlet edge through the nearest neighbor network of grains. We then define the clogged state as the first frame wherein the final arch grains fall within $0.3$~mm $\sim 0.035 d_L$ of their final positions, which is our tracking precision. 

Following prior work~\cite{thomas_fraction_2015} we expect that the `clogging' state will be one sampling time $\tau = d/v_{outlet} \approx 3$~frames before the clogged state. However, because of the intermittent nature of flow, we find that $3$~frames before the clogged state is occasionally a nearly-identical configuration. An example of this scenario is shown in row 5 of Fig.~\ref{fig:StateDef}. Here a meta-stable arch is created that temporarily halts flow but ultimately fails without external forcing. Therefore to create a consistent label we define the clogging state as follows. We conceptualize the final arch grains like a fence, fully contacting in the clogged case and only just `closing' in the clogging case. The clogging state is thus the earliest configuration where the total sum of spaces between all final arch grains is not enough to admit another (small) grain to pass between them. We define flowing states as any state prior to \textit{Clogging}, but exclude states that occur within three sampling times (nine frames) of the clogging state from our classification analysis. We also exclude momentarily static `flowing' states (meta-stable arches) that occur in the middle of flow, as in this work we are interested primarily in distinguishing between actively flowing and permanently clogged states.

\section{SVM Cost Minimization}\label{SVMFeatureContribution}
The cost function $\mathcal{L}$ for the soft-margin linear SVM we have used in this work contains two parts,
\begin{equation}
    \mathcal{L}(\mathcal{C}_i,L_i) = \underbrace{\sum_{i} \text{max}[0,1-\mathcal{C}_{i}L_{i}]}_{\text{hinge loss}}
    + \underbrace{\lambda \sum_{j} \theta_{j} }_{\text{regularization}}
\end{equation}
where $i$ and $j$ are indices of training set microstates and of tunable weights $\theta$ respectively, and $\lambda$ is a regularization coefficient. The hinge loss term penalizes data for which the output is incorrect or incomplete, \textit{i.e.} the product of the Clogginess value $\mathcal{C}_{i}$ and the label $L_{i} = \pm 1$ is below $1$. The regularization term promotes sparse weights, while also preventing the system from simply spreading out its output values to lower the cost (as doubling all $\theta$ doubles all $\mathcal{C}$.) As a result, $\lambda$ controls the trade off between the cost of misclassification and the spreading of $\mathcal{C}$ values.

To prevent features of different scales, \textit{e.g.} the inner most and outermost rings of the Behler-Parrinello functions in Figure~\ref{fig:SVM_BP}a, from unevenly affecting the decision function, we standardize the feature functions to have a mean of 0 and standard deviation of 1. This method addresses mismatched scales, but assumes a roughly Gaussian distribution of values for each feature. If this isn't the case, it may incidentally inflate or contract the available values of a feature to the cost minimization. For this reason we avoid standardizing our circular window functions; these functions are already area fractions, bounded between 0 and 1 and of generally similar scale.

\begin{figure*}[t!]
    \centering
    \includegraphics[width=15cm]{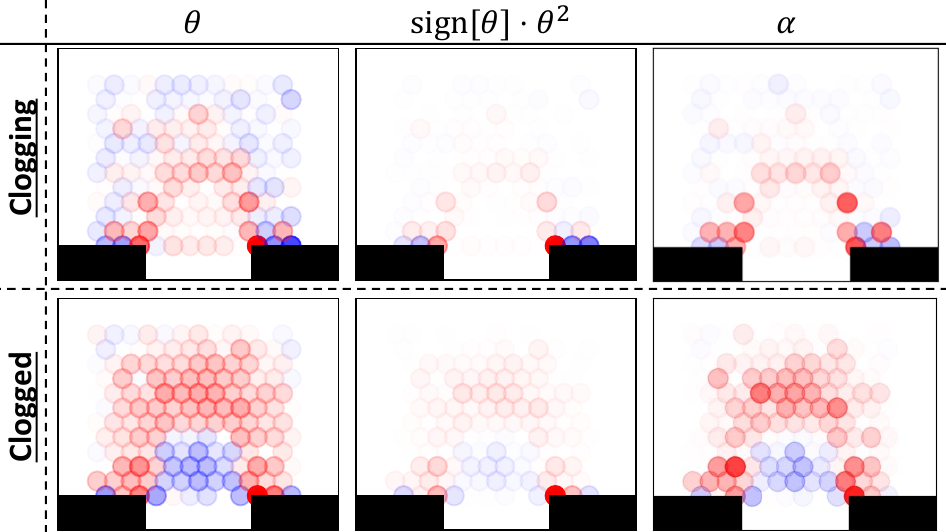}
    
    \caption{Comparison of linear and signed squared feature weights, $\theta_j$ and $\theta_j^{2}$, with feature significance, $\alpha_{j} = \mathrm{sign(\theta_{j})} \text{var}[\theta_j \times G^{DG}_j(\Omega)]_{\text{training set}}$, for clogging- and clogged-trained models. For both training sets, the three visualizations show different features. Color scales are logarithmic, as in Fig.~\ref{fig:SVM_Baseline}, with color map limits determined by the maximum absolute value of each plot independently. We note that varying window size between one quarter and twice the area, $d_{\mathrm{wind}} = [.5 d_{\mathrm{S}},\sqrt{2} d_{\mathrm{S}}]$, yielded qualitatively identical pictures.}
    \label{fig:WeightsContribsExample}
\end{figure*}

In either case, the weights fail to capture the full contribution of a feature dimension to the final result. This motivated the construction of the feature significance, $\alpha$. We observe the difference in Figure~\ref{fig:WeightsContribsExample} by plotting the linear and signed squared feature weights, $\theta$ and $\theta^{2}$, as well as feature significance, $\alpha$, for clogging- and clogged-trained models. The colormap is logarithmic, as in Fig.~\ref{fig:SVM_Baseline}, however each plot is normalized independently as the scales are different.

Raw weights $\theta$ seem to show large range features with the most apparent points of interest being the transitions in the sign of $\theta$ rather than individual areas of import. The squared feature weights isolate only the features of largest weight, primarily along the bottom of the hopper. However, we note that these windows are also partially occluded; the maximum area overlap is lower than windows higher in the system. Thus is the larger weight given by the SVM compensating for the smaller values, or is there additional value of the information in that region? This was the motivating question for feature significance. Notably, the the signed squared feature weights is a special case of the feature significance, wherein all features have the same variance. Because the two values are not the same, we can immediately see there was structural information absent when looking only at the weights themselves. In particular, some of the large range information returns, and we can confidently say the regions along the bottom edge of the hopper are more significant than those in the bulk.

\section{Alternate Analyses}\label{PastWork}
In addition to the work presented above, we investigated a variety of alternative analyses that were less accurate or interpretable in their results. We first present the Behler-Parrinello inspired functions, which showed agreement with theoretical predictions of where information is located spatially, but gave no additional insight. We then describe alternative structure functions that were not as effective, and finally entirely different machine learning models that failed to improve on the accuracy presented above. 
\begin{figure*}
    \centering
    \includegraphics[width=17cm]{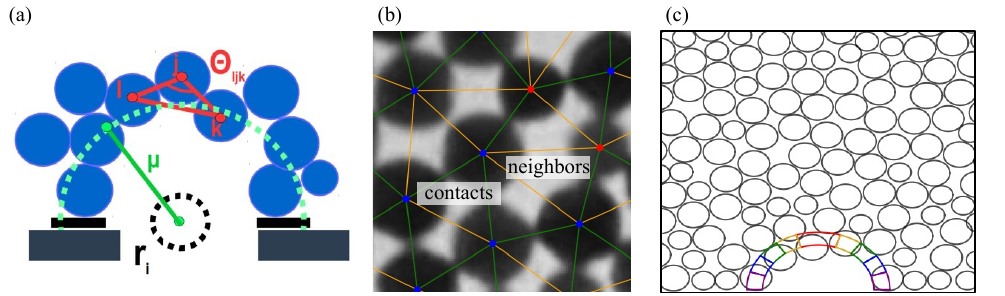}
    \caption{Three independent methods of parameterizing configurations that performed worse than the methods presented above. (a) A schematic for calculating modified versions of the Behler-Parrinello angular distribution functions. These functions bin grains by their distance from the center of the outlet, $\mu$, and only calculate the angles between grains that are in or near contact. (b) An example of the Delaunay triangulation used to define local neighbor network metrics. Green connections are considered near contact, while orange connections are only neighbors. The sum of triangular areas for which grain i are a vertex are considered the free area of that grain. (c) An example radius of hopper symmetric functions. Area overlaps are calculated for each window, and windows symmetric about the vertical axis bisecting the outlet (windows of the same color) are summed together. }
    \label{fig:FailedSFs}
\end{figure*}

\subsection{Behler-Parrinello Functions}
\begin{figure*}[t!]
\includegraphics[width=17.5cm]{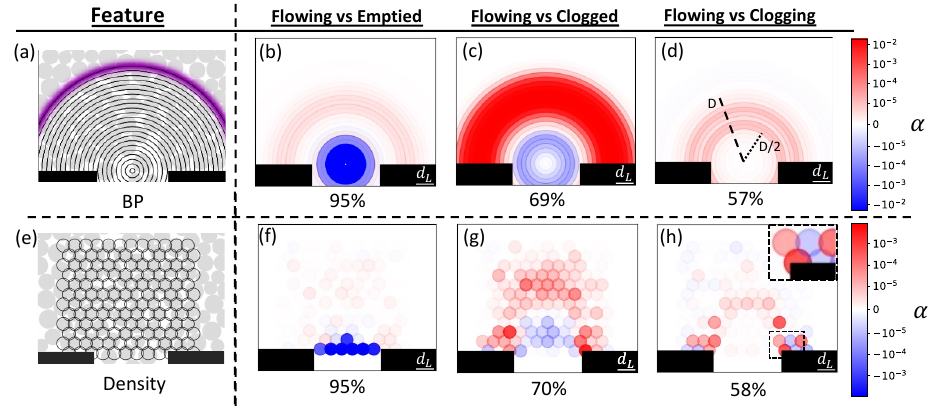}
\caption{ Example structure function windows (a) and feature significance $\alpha_{j}$ (signed variance of feature contributions across training data) (b-d) for the Behler-Parrinello inspired features for each of the three binary classification tasks: Clogging, Clogged, or Emptied vs Flowing. Features in blue (red) indicate additional (reduced) density in that region makes a Flowing prediction more likely, and the intensity of the color indicates the magnitude of the effect. Feature significance is narrowly restricted between two length scales determined by the outlet size, $[D/2,D]$. The density grid functions (e) and weights (f-h) are shown again for comparison. }
\label{fig:SVM_BP}
\end{figure*}

An alternative set of successful structure functions are written as:
\begin{equation}
    G^\mathrm{BP}_{n} = \sum_{\text{grains}} e^{-(|\vec r |-\mu_{n})^{2}/L^{2}} 
    \quad , \quad G^\mathrm{BP}_0 = 1
\end{equation}
where $G^\mathrm{BP}_n$ calculates the radial density of grain centers $\vec r$ over Gaussian rings of constant width $L$, and $G^\mathrm{BP}_0$ again gives the system an adjustable offset. Each $G^\mathrm{BP}_n$ calculated from a microstate corresponds to a density at distance $\mu_n$ relative to the center of the outlet. $\mu_n$ values are spaced evenly by $\Delta \mu_n$. These functions are adapted from Behler and Parrinello~\cite{behler_generalized_2007} (hence $BP$), have been useful in identifying structural properties of other particulate systems~\cite{cubuk_identifying_2015,harrington_machine_2019}, and are pictured in Fig. \ref{fig:SVM_BP}a. Much like the $DG$ structure functions, varying the hyperparameters for these $BP$ functions ($\Delta \mu_n$ and $L$) over the range $[0.5d_{S},1.0d_{L}]$ had no impact on final accuracy.

These Behler-Parrinello-inspired features ($G^\mathrm{BP}$) are calculated relative to the hopper geometry, and thus are effective for identifying system scale features of clogging. We show the feature signficance for \textit{Flowing} vs \textit{Clogging}, \textit{Clogged}, and \textit{Emptied} states in Fig.~\ref{fig:SVM_BP}(b-d) respectively. We see two clear length scales emerge, $r_\mathrm{ring} = D/2$ and $r_\mathrm{ring} = D$. Our machine-learned solutions find no relevant contributions outside this larger radius, consistent with the modelling of Thomas and Durian~\cite{thomas_fraction_2015} of the scale of the region relevant for clogging, $\propto D^2$ in the case of a quasi-2D hopper. Note that we have purposely identified clogging states in such a way that they cannot be identified by the lack of grains within the outlet, as this would be trivial. However, in predicting clogged and emptied states (Fig.~\ref{fig:SVM_Baseline}b and c) this becomes a useful metric, as denoted by the small blue rings. 

\subsection{Alternate Structure Functions}
During the linear SVM analysis, we experimented with a wide array of structure functions. We describe three additional structure function families to those mentioned above, as they either performed the same or worse on the data set without offering any additional insight.

\subsubsection{Angular Distribution Functions}
We also attempted an extended version of the Behler-Parrinello structure functions. Above, we adapted the density functions for use relative to the hopper geometry rather than surrounding a single grain. In the original description by Behler and Parrinello~\cite{behler_generalized_2007} and the later applications to glassy systems~\cite{schoenholz_structural_2016} there are also angular functions of the form:
\begin{equation*}
    R_{ijk} = e^{((r_{ij}^{2}+r_{jk}^{2}+r_{ki}^{2})/\xi^{2}}
\end{equation*}
\begin{equation*}
    \Phi_{ijk} =(1+\lambda \mathrm{cos}(\Theta_{ijk}))^{\zeta}
\end{equation*}
\begin{equation}
    \psi_{i} (\xi,\lambda,\zeta) = \sum_{i} \sum_{j} \sum_{k}  R_{ijk} \Phi_{ijk} \label{Eq:BPAngles}
\end{equation}
Where $\xi,\lambda$ and $\zeta$ are all varied over different ranges. These functions, as indicated by the subscript on $\psi_{i}$, would be calculated relative to grain i, with $\Theta_{ijk}$ representing the angle subtended by the vectors $\vec{r_{ij}}$ and $\vec{r_{ik}}$. All three grains must be within some mutual length scale set by $\xi$. As we noted with the grain densities, clogging is a function of an entire state rather than a single grain and so these functions needed to be calculated relative to the geometry, in particular we chose the center of the outlet. Thus we added the same radial binning as the density functions above as a prefactor, $g_{i}$:
\begin{equation*}
    g_{i} = e^{((r_{i}-\mu)/L)^{2}}
\end{equation*}
\begin{equation}
    \psi_{\mu} (L,\xi,\lambda,\zeta) = \sum_{i} \sum_{j} \sum_{k} g_{i} g_{j} g_{k} R_{ijk} \Phi_{ijk} \label{Eq:OurAngles}
\end{equation}
Thus our angular functions are now binned by radial distance $\mu$, as in in Fig.~\ref{fig:SVM_BP}a, and summed over each grain $(i,j,k)$. Instead of varying the mutual length scale between grains, $\xi$, we were primarily interested in three grains near contact and the angular distribution of those structures. Thus we set $\xi=d_{L}$ to keep the summation dominated by neighbor interactions. Unfortunately, these functions failed in a variety of ways. Not only were they difficult to interpret physically, but also were given little weight by the SVM. In fact the accuracy remained unchanged when these angular functions were removed. It is possible that binning these functions by distance averaged the information we were looking for over so many permutations as to no longer be discriminating.

\subsubsection{Local Neighbor Network Functions}
The second family of structure functions we constructed described the local network of contacts and area available to each grain. We calculated the Delaunay triangulation of the grains to approximate the edges of the neighbor network. Edges that were larger than $2d_{L}$ were pruned for being too far to be neighbors. Edges within 10\% of the sum of the connecting grains, $\vec{r_{ij}} \leq 1.1*(r_{i}+r_{j}$ are considered `near contact'. Finally, the Voronoi tesselation is used to calculate the free area surrounding each grain. The number of neighbors, number of contacts, and free area are calculated for each grain and then binned by distance. These functions again yielded no difference over randomness, and we suspect the binning the information by distance may have averaged the discrepant information away.

\subsubsection{Hopper Symmetric Area Overlap Functions}
The third family of structure functions can be thought of as a subset of the area overlap functions presented above. We constructed functions that reflected the symmetry of the hopper geometry; area overlaps were calculated over arch shaped windows instead of circular ones. In addition, windows placed symmetrically about the vertical line that bisects the outlet were summed together. These functions were found to perform worse than the grid of circular windows we present in this work. We expect this is due to the a priori summation of symmetrically placed windows; we expect the final weights to be symmetric about the outlet but individual states are not. Thus, like the other structure function families that failed, the discriminating asymmetries were averaged out.

\subsection{Inclusion of Velocities}

Previous work has suggested that grain momenta do not affect whether a clog will form from a given positional microstate~\cite{koivisto_effect_2017}. To test this hypothesis we add velocity information to our SVM analysis. We calculate the velocities of grains in both $\hat x$ and $\hat y$, weighted by the area fraction $G^\mathrm{DG}_{n}$. We do this also by size, resulting in a further tripling of our structure functions from the area-only case. Consistent with previous work, we find minimal benefit of including velocities when predicting \textit{Clogging} states, as shown in Table~\ref{tab:Accuracies}. 

\subsection{CNN Reconstruction}

We also identify states using a Convolutional Neural Network (CNN). In this method the function $f_{\vec \theta}$ has a complex nonlinear form and receives an \textit{image} of the microstate $I(\Omega_i)$ as its input. To speed processing, we both restricted the field of view of this image to the same region as the SVM functions and represented grains as 2x2 pixel regions of color with Gaussian weighted intensity. Grain sizes were represented by independent color channels making each reconstruction a 3-channel image 32x32 pixels large. Our peak validation accuracies, reported in Table~\ref{tab:Accuracies}, use this representation in a 830,000~parameter, 35~layer convolutional neural network. Alternate reconstruction techniques and network architectures were implemented and are described in the following section, however most were within $2\%$ of the values in Table~\ref{tab:Accuracies} and thus we do not believe the differences are significant.

\subsection{Alternate CNN Reconstruction}
\begin{figure}[t!]
    \centering
    \includegraphics{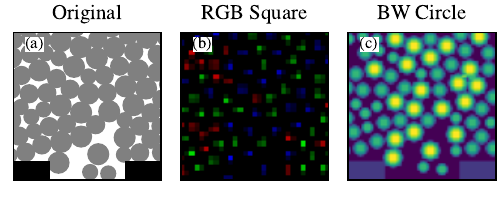}
    \caption{Example reconstructions used in the CNN analysis. The original reconstruction (a), 2x2 Gaussian squares (b), and the direct black and white downsampling (c) are all shown for the same microstate with the same field of view as the linear SVM structure functions.}
    \label{fig:CNNRecon}
\end{figure}

In addition to the Gaussian regions we utilized in the CNN above, we also attempted a direct downsampling of the full size black and white reconstruction. This downsampling was 64x64 pixels large, twice the size of the RGB reconstruction, with each grain being drawn to scale using the midpoint circle algorithm. Circles were drawn with uniform brightness, as well as linearly increasing brightness towards the center, with the latter having a higher validation accuracy. A comparison of the different reconstructions can be found in Figure~\ref{fig:CNNRecon}. We also implemented a variety of methods for representing the bottom of the hopper in both reconstructions, however this never resulted in an observable difference in validation accuracy.

We also implemented several CNN architectures. The number of convolutional layers and the number of filters in each layer were independently varied over an order of magnitude. The size of the kernel as well as the frequency of skip connections and max/average pool layers, presence and severity of dropout, the number of nodes in each dense layer was varied between $[8,128]$ and the number of layers varied between $[1,10]$. 

\subsection{Configuration Time and Support Vector Regression}
As we note above, clogging is a Poisson process wherein the flow randomly samples states at a constant rate until one that forms a clog is found. We also note that this is true macroscopically, but for any given flow the flowrate fluctuates substantially around the average value, which makes the definition for the clogging state challenging. A previous definition for identifying unique configurations involved a quantity we defined as configuration time, $t_{\mathrm{config}}$. The configuration time specified how long from a given state until the 2D autocorrelation function decreases by $1/e$ towards its steady state value. The median value for configuration time matched the sampling time of the system, but was also dynamic to capture any conditions that may alter how long a given configuration is present.

One benefit of configuration time being a dynamic variable attached to each state is it allowed for a different method of linear Support Vector Machine, Support Vector Regression. We would still characterize our configurations the same as outlined above, in particular using the Behler-Parrinello functions. However, instead of labelling states by class and finding the hyperplane (that is the equation for  $\mathcal{C}=0$) that best separates two classes, we simply fit a variable, $t_{\mathrm{config}}$ to a line. This had several possible benefits for our system. Because clogging is a rare event in the flow, collecting data naturally yielded approximately 100 flowing states for every clogging state. We could also include meta-stable arches, excluded in all the above analysis, as they would simply be more points near clogging to fit.

However, our core question is still classification, so we broke down this approach into two parts. First, we performed linear regression against the configuration time of the next state; flowing states lead to more flowing states with finite configuration times but clogging states lead to the clogged state with functionally infinite configuration time. For our purposes we set this as a finite number 2 orders of magnitude larger than our largest times. We also tried regressing to inverse configuration time, $1/t_{\mathrm{config}}$, so the arbitrary choice $t_{\mathrm{config}}$ would converge to 0 instead of diverge. 

In either case, we weren't interested in the final regressed values as the accuracy of a linear regression in this system struggled to give accurate results. However, following the work of Rocks, Ridout and Liu~\cite{rocks_learning-based_2021}, we hoped the hierarchy of values would be preserved and could be used for classification. Effectively we wish to calculate the cumulative distribution of regressed $t_{\mathrm{config}}$ or $1/t_{\mathrm{config}}$ values, wherein we expect the clogging state to be an extrema, either the greatest (for $t_{\mathrm{config}}$) or the least (for $1/t_{\mathrm{config}}$). We found the median percentile of clogging states was only slight different than 50 (60th percentile for $t_{\mathrm{config}}$, 34th percentile for $1/t_{\mathrm{config}}$). There were a few factors that could contribute to this muddled result. First, and most prominently, this was a much smaller data set of $\approx 1500$~flows using metal bearing balls rather than quasi-2D grains. There were concerns of the smaller radius bearing balls having a quasi-3D effect depending on whether they were at the forward or rear edge of the hopper. But ultimately this method involves much of the same tools as the SVM's presented above, but through the lens of a newly quantified quantity that may not be possible to predict. If a different, directly measured quantity of the system could be used, perhaps this method might prove a useful way to extrapolate from the large amount of unused flow data.

\subsection{Graph Neural Networks} \label{GraphNets}
Following the work in Ref.~\cite{battaglia_relational_2018}, Graph Neural Networks (GNN's) offer a seemingly-natural approach to the problem of clogging. The node and edge structure are well suited to describe the contact network we believe to be crucial to the understanding of how clogs form. While we experimented with many representations of which data to include, the most complete was as follows. Most nodes represents an individual grain and encode the properties for that grain: $\{x, y, r, d_{\mathrm{left~outlet}}, d_{\mathrm{right~outlet}},v_{\mathrm{x}},v_\mathrm{{y}}\}$. In addition to the nodes representing grains, we included two nodes of a different size to represent the edges of the outlet, and a third node to represent the bottom edge of the hopper.
 
To identify edges, a Delaunay triangulation was used to identify spatially-neighboring nodes. Edges of greater than 2~$d_{\mathrm{L}}$ were pruned for being too far to consider `near contact'. This neighbor network then formed a bidirected graph wherein each neighbor pair was represented twice as two directed edges in opposite directions. Additional edges were then added for all the grains in the neighborhood of the outlet edges, and any grain in contact with the hopper bottom, with edges only going in the direction of the hopper geometry out to the grains. Each edge encoded the absolute, signed displacement of from source to target (so edges of opposite direction would have opposite sign) as well as the gaps between grains, $\mathrm{gap}_{ij} = \mathrm{displacement}_{ij}-r_{i}-r_{j}$. As noted in the discussion above, we suspect these gaps or voids may be critical to understanding clogging at the microstate level. Finally, for training tasks where we attempted to identify flowing from clogging as a global state, a global variable was also included to represent this binary classification.

We attempted multiple training targets, listed in Table~\ref{tab:GNN}. The core question presented in this work, flowing vs clogging, received similar results in the GNN as in the linear SVM, approximately $57\%$ classification accuracy when training flowing vs clogging, and $99\%$ accuracy for flowing vs clogged. Given this is worse than the accuracies we find with the Convolutional Neural Nets, we are unsure where the discrepancy lies. We speculate that a different data representation may improve performance as perhaps a symmetric representation of all contacts belies the fundamental asymmetries that make clogging an interesting phenomena in the first place.
 
Next we investigated a variety of per-grain training protocols. First, we trained for whether individual grains were flowing or clogging, as was originally outlined by Thomas and Durian~\cite{thomas_fraction_2015}. Unfortunately, this presents difficulties since we expect many grains that are individually in a clogging state will continue flowing, as clogging is a collective phenomenon when all the grains the region of the outlet are clogging. When we labeled nodes as flowing or clogging by the global state, we are accidentally mislabeling many clogging grains as flowing. Whether because of this approximation or other limitations, this method yielded no increase in accuracy from the baseline $50\%$ expected for a binary classification task. 
 
We also attempted to predict node-based dynamic quantities that could be proxies for clogging. We attempted to predict binned grain velocities; we separated the velocities present in our system into 10 bins and attempted a 10 state classification task. We found a $22\%$ accuracy, which is an improvement over the baseline $10\%$. Approximately $50\%$ of grain velocities were within one bin of the correct label, so one might consider a more coarse-grained binning protocol, but ultimately it was unclear what information could be gleaned from velocities that uncertain. Finally we attempted to predict the sign of a grain's acceleration, whether it was slowing down or speeding up, which we were able to do with $66\%$ accuracy. Similar to the velocities, given the relatively small increase above baseline $50\%$, we didn't think this warranted further investigation for what was ultimately a proxy metric for the core investigation.
 
\begin{table}
\caption{\label{tab:GNN} The different training targets and their test set accuracies using Graph Neural Networks (GNNs)}
\begin{ruledtabular}
\begin{tabular}{cc}
 Training Goal & Test Set Accuracy\\ \hline
 Flowing vs Clogging (Global) & 57$\%$  \\
 Flowing vs Emptied Arch (Global) & 99$\%$  \\
 Flowing vs Clogging (Nodes) & 50\% \\
 Binned Grain Velocities (Nodes) & 22\% \\
 Sign of Grain Acceleration (Nodes) & 66\% \\
\end{tabular}
\end{ruledtabular}
\end{table}

\section{Predicting Individual Clog Formation}\label{Clogginess}
\begin{figure}[t!]
\includegraphics[width=8.5cm]{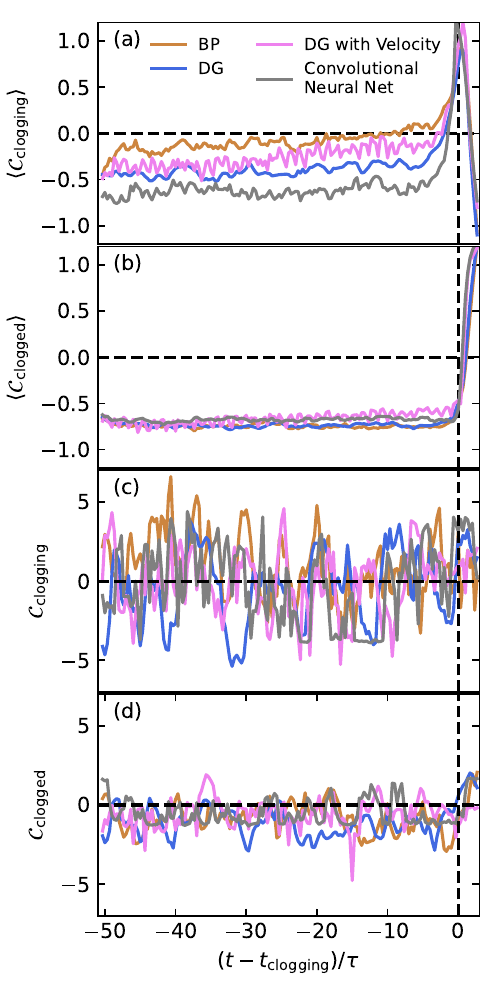}
\caption{\label{fig:Clogginess} Signed distance from the hyperplane, Clogginess, versus flow time for clogging- and clogged-trained models aligned by the clogging frame. (a,b) Clogginess values are averaged over 2000 flow events excluded from the training set for the (a) clogging-trained and (b) clogged-trained models. The standard deviation, not shown, is $\approx 5$ for each curve. (c,d) Clogginess versus flow time for a single example flow is plotted for the (c) clogging-trained and (d) clogged-trained models. In all plots the dashed black lines are plotted at 0 values as a guide to the eye.
}
\end{figure}

We next investigate the ability of the above methods to predict a single clog event as a flow progresses. That is, we observe the state prediction from each flow microstate over time $\mathcal{C}(t)$. We do this for the clogging- and clogged-trained models, conceptualizing the latter as an extreme case of our core investigation. In previous work on particulate systems~\cite{cubuk_identifying_2015}, a machine-learned quantity (`Softness') was used to predict structural changes. Thus we refer to any of our machine-learned outputs $\mathcal{C}(t)$ as `Clogginess,' though unlike the above referenced work we use both linear and nonlinear methods for manifesting this prediction. We normalize the Clogginess values from each model to have a range of 2 as the overall magnitude of this range is arbitrary. In Fig.~\ref{fig:Clogginess}, we plot the average Clogginess value for each model vs time over 2000 flows, aligned by the clogging frame. 

These results demonstrate a clear signal: the states before $t_{\mathrm{clogging}}$ have a consistently negative average Clogginess value, and states of $t \geq t_{\mathrm{clogging}}$ have positive average Clogginess, with a rapid transition in between. This transition occurs in one to four sampling times $\tau$, in qualitative agreement with the Poissonian dynamics of previous work\cite{thomas_fraction_2015}. Of note, the models trained for flowing vs clogged (Fig.~\ref{fig:Clogginess}b) have similar average Clogginess curves but very different test-set accuracy, stemming from their variation in fluctuation magnitude (Fig.~\ref{fig:Clogginess}d). The models trained for flowing vs clogging (Fig.~\ref{fig:Clogginess}a) have similar test-set accuracies, while their average curves differ significantly. We are at present unsure if the distinction in average curves is a function of the choice of structure function, choice of normalization, or something else entirely, but the discrepancy motivates a more direct examination of the structure function as is shown later in this paper.

These results show that Clogginess is -on average- predictive of the onset of a clog, but does not translate to the scale of a single flow event. The Clogginess values for a single flow, as in Fig.~\ref{fig:Clogginess}c, lose the characteristic features outlined above. For each case, the variance from state to state is an order of magnitude higher than the average signal, making individual predictions not useful.

\section{Insensitivity to Particle Tracking Uncertainties}\label{AddedNoise}
In static stable packings grain-grain contact forces play a major role, and hence there is a significant different between grains that are in contact and those that are infinitesimally away from contact.  So it is natural to wonder if inevitable limitations on experimental precision are the source of poor classification accuracies. To investigate this we repeated our density grid SVM analysis on the same data but with artificial noise added to each grain position in each microstate. Specifically, we add Gaussian noise to the $x$ and $y$ positions of each grain with zero mean and different standard deviations $\sigma_{0}$. We report the prediction accuracy for each noise level with an effective noise strength $\sigma_{\text{noise}} \equiv \sqrt{\sigma^2_0 + \sigma^2_{\text{tracking}}}$. $\sigma_{\text{tracking}} = 0.14~\mathrm{mm} \approx 0.01d_L$ represents the measured uncertainty in our grain tracking, calculated from the standard deviation in grain positions filmed under static conditions.

\begin{figure*}[t!]
\includegraphics[width=17cm]{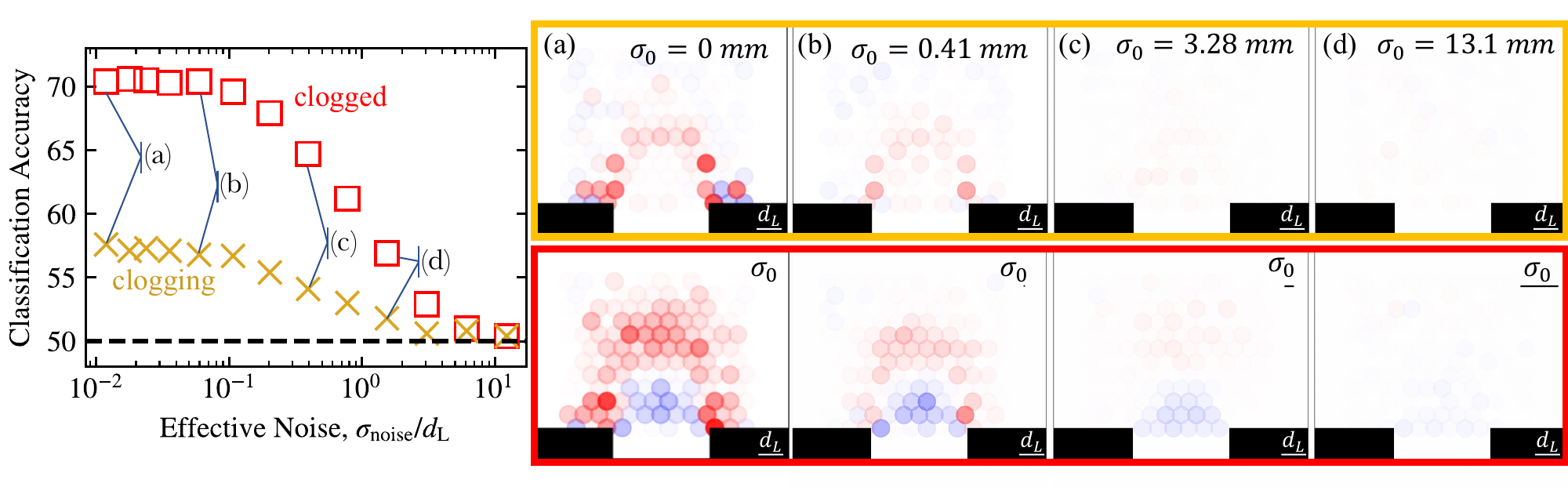}
\caption{\label{fig:Sup_AddNoise} Effect of added noise to grain positions on SVM decision accuracy. (Left) SVM test set accuracy versus effective noise, $\sigma_{\text{noise}} \equiv \sqrt{\sigma^2_0 + \sigma^2_{\text{tracking}}}$ for clogging-trained (yellow circles) and clogged-trained (red squares) models. (a-d) Visualizations of feature significance $\alpha$ maps for increasing added noise $\sigma_0$. As we increase the injected noise from baseline (a), we initially see no loss in test set accuracy and feature significance retains its overall structure but loses intensity (b). If the total noise exceeds $\approx 0.1d_{\mathrm{L}}$ the classification begins to lose accuracy. The largest magnitude features, those near the edges of the outlet, disappear entirely (c) and as the noise is further increased all features lose significance (d).}
\end{figure*}

We find that with noise that does not exceed the experimental uncertainty by over an order of magnitude ($\sigma_{\text{noise}} \leq 0.1d_L$), the accuracy does not substantially decrease, as shown in Figure~\ref{fig:Sup_AddNoise}. Over this range, we see the feature weights in the cornerstone region reduce in intensity as the noise increases, but remain the same in sign and position, as shown in Fig.~\ref{fig:Sup_AddNoise}a to b. This suggests that the same features remain available, but are reduced in salience (as expected). This also indicates that our experimental uncertainty is likely not a limiting factor for the observed classification accuracies. For larger noise, $\sigma_{\text{noise}} > 0.1d_L$, the accuracy degrades until it completely eliminates any relevant features and the system cannot predict above random chance ($50\%$). Comparing the feature weights before (Fig.~\ref{fig:Sup_AddNoise}b) and after (Fig.~\ref{fig:Sup_AddNoise}c) the change in accuracy, we see that the prediction moves from small-scale grain positions at the outlet to (faint) bulk densities. Overall, the addition of noise strengthens the conclusion that the availability of potential cornerstones is critical to differentiating clogging vs flowing, and the model is sensitive to the cornerstone locations to a length scale near $0.1d_{\mathrm{L}}$.

\section{Possible Future Directions}\label{FutureWork}
Given the wide range of possible machine learning algorithms, we believe there are still many avenues to continue the investigation of clogging at scale. Using the knowledge we gained in the development of the SVM's and CNN's utilized in this paper, we believe one or more of the following analyses may be fruitful.

\subsection{Autoencoder for determining minimal space representation of data}
This work presents the first models we are aware of that classify whether a state is going to continue flowing or cause a clog. As such, it is difficult to identify potential bottlenecks to our accuracy. In particular for the nonlinear methods, are we limited by the data quantity, variable representation, or the network complexity? One possibility for restricting the phase space of possible analyses is an autoencoder. An autoencoder takes in a representation of a state and attempts to reproduce that state after being forced into a dimensionally-reduced space. An autoencoder that successfully reconstructs the input state could inform the minimal neural network architecture required to describe a state in a CNN, for example. Additionally, the dimensionally-reduced space, or latent space, that input data is transformed into can characterize the dimensionality of information included in the state. How this latent space changes for controlled changes in the input information could inspire future choices of state description.

\subsection{CNN using Gaussian density kernels of different size}
In this work we found nonlinear methods presented a modest improvement over the more interpretable linear SVM for classifying flowing and clogging states. However, we acknowledge our exploration of non-linear methods is nowhere near exhaustive. One avenue we would explore in the future is alternate methods of data representation for the CNN. We tried two methods: downsampling the full reconstruction and an abstract reconstruction where grains of different sizes were on different layers. In both cases, the CNN only has access to local length scales set by the size of the convolutional layers. An alternate reconstruction method would using Gaussian kernels of different sizes on the full reconstruction as the layers of a single image. This might make features of different sizes more accessible to the CNN.

\subsection{Alternate Structure Functions: Gaps between grains, Consistent Final Arch Placement}
We believe the contact network throughout the hopper could be invaluable to understanding how stable arches form. Unfortunately, the contacts in our system are frictional and our grains are not photoelastic, meaning whether grains are in contact and whether those contacts are loaded is not information we can access. That said, one could consider directly parameterizing the gaps between grains as an approach to this subject. In combination with parameterizing the contact network, one could consider spatially localizing parameters by the `arch' field of view. One explanation for why the cornerstones are highlighted by the linear SVM, rather than keystones, is the keystone for any given arch may exist in a wide variety of locations, whereas the cornerstones were found to be consistently located. For a linear combination of area overlap functions, a signature of clogging appearing in different locations is impossible to identify. If a reference point was chosen that should always include the final arch, such as the shortest path from one cornerstone to the other, then perhaps the linear SVM may still succeed.

\bibliography{references}

\begin{thebibliography}{39}%
\makeatletter
\providecommand \@ifxundefined [1]{%
 \@ifx{#1\undefined}
}%
\providecommand \@ifnum [1]{%
 \ifnum #1\expandafter \@firstoftwo
 \else \expandafter \@secondoftwo
 \fi
}%
\providecommand \@ifx [1]{%
 \ifx #1\expandafter \@firstoftwo
 \else \expandafter \@secondoftwo
 \fi
}%
\providecommand \natexlab [1]{#1}%
\providecommand \enquote  [1]{``#1''}%
\providecommand \bibnamefont  [1]{#1}%
\providecommand \bibfnamefont [1]{#1}%
\providecommand \citenamefont [1]{#1}%
\providecommand \href@noop [0]{\@secondoftwo}%
\providecommand \href [0]{\begingroup \@sanitize@url \@href}%
\providecommand \@href[1]{\@@startlink{#1}\@@href}%
\providecommand \@@href[1]{\endgroup#1\@@endlink}%
\providecommand \@sanitize@url [0]{\catcode `\\12\catcode `\$12\catcode
  `\&12\catcode `\#12\catcode `\^12\catcode `\_12\catcode `\%12\relax}%
\providecommand \@@startlink[1]{}%
\providecommand \@@endlink[0]{}%
\providecommand \url  [0]{\begingroup\@sanitize@url \@url }%
\providecommand \@url [1]{\endgroup\@href {#1}{\urlprefix }}%
\providecommand \urlprefix  [0]{URL }%
\providecommand \Eprint [0]{\href }%
\providecommand \doibase [0]{https://doi.org/}%
\providecommand \selectlanguage [0]{\@gobble}%
\providecommand \bibinfo  [0]{\@secondoftwo}%
\providecommand \bibfield  [0]{\@secondoftwo}%
\providecommand \translation [1]{[#1]}%
\providecommand \BibitemOpen [0]{}%
\providecommand \bibitemStop [0]{}%
\providecommand \bibitemNoStop [0]{.\EOS\space}%
\providecommand \EOS [0]{\spacefactor3000\relax}%
\providecommand \BibitemShut  [1]{\csname bibitem#1\endcsname}%
\let\auto@bib@innerbib\@empty
\bibitem [{\citenamefont {Nedderman}\ \emph {et~al.}(1982)\citenamefont
  {Nedderman}, \citenamefont {Tuzun}, \citenamefont {Savage},\ and\
  \citenamefont {Houlsby}}]{nedderman_flow_1982}%
  \BibitemOpen
  \bibfield  {author} {\bibinfo {author} {\bibfnamefont {R.~M.}\ \bibnamefont
  {Nedderman}}, \bibinfo {author} {\bibfnamefont {U.}~\bibnamefont {Tuzun}},
  \bibinfo {author} {\bibfnamefont {S.~B.}\ \bibnamefont {Savage}},\ and\
  \bibinfo {author} {\bibfnamefont {G.~T.}\ \bibnamefont {Houlsby}},\
  }\bibfield  {title} {\bibinfo {title} {The flow of granular materials-1:
  {{Discharge}} rates from hoppers},\ }\href@noop {} {\bibfield  {journal}
  {\bibinfo  {journal} {Chem. Eng. Sci.}\ }\textbf {\bibinfo {volume} {37}},\
  \bibinfo {pages} {1597} (\bibinfo {year} {1982})}\BibitemShut {NoStop}%
\bibitem [{\citenamefont {Helbing}\ \emph {et~al.}(2000)\citenamefont
  {Helbing}, \citenamefont {Farkas},\ and\ \citenamefont
  {Vicsek}}]{helbing_simulating_2000}%
  \BibitemOpen
  \bibfield  {author} {\bibinfo {author} {\bibfnamefont {D.}~\bibnamefont
  {Helbing}}, \bibinfo {author} {\bibfnamefont {I.}~\bibnamefont {Farkas}},\
  and\ \bibinfo {author} {\bibfnamefont {T.}~\bibnamefont {Vicsek}},\
  }\bibfield  {title} {\bibinfo {title} {Simulating dynamical features of
  escape panic.},\ }\href@noop {} {\bibfield  {journal} {\bibinfo  {journal}
  {Nature}\ }\textbf {\bibinfo {volume} {407}},\ \bibinfo {pages} {487}
  (\bibinfo {year} {2000})}\BibitemShut {NoStop}%
\bibitem [{\citenamefont {Olson~Reichhardt}\ and\ \citenamefont
  {Reichhardt}(2013)}]{olson_reichhardt_vortex_2013}%
  \BibitemOpen
  \bibfield  {author} {\bibinfo {author} {\bibfnamefont {C.~J.}\ \bibnamefont
  {Olson~Reichhardt}}\ and\ \bibinfo {author} {\bibfnamefont {C.}~\bibnamefont
  {Reichhardt}},\ }\bibfield  {title} {\bibinfo {title} {Vortex {{Clogging}},
  {{Commensuration}}, and {{Diodes}} in {{Asymmetric Constriction Arrays}}},\
  }\href@noop {} {\bibfield  {journal} {\bibinfo  {journal} {Journal of
  Superconductivity and Novel Magnetism}\ }\textbf {\bibinfo {volume} {26}},\
  \bibinfo {pages} {2005} (\bibinfo {year} {2013})}\BibitemShut {NoStop}%
\bibitem [{\citenamefont {Zuriguel}\ \emph {et~al.}(2014)\citenamefont
  {Zuriguel}, \citenamefont {Parisi}, \citenamefont {Hidalgo}, \citenamefont
  {Lozano}, \citenamefont {Janda}, \citenamefont {Gago}, \citenamefont
  {Peralta}, \citenamefont {Ferrer}, \citenamefont {Pugnaloni}, \citenamefont
  {Cl{\'e}ment}, \citenamefont {Maza}, \citenamefont {Pagonabarraga},\ and\
  \citenamefont {Garcimart{\'i}n}}]{zuriguel_clogging_2014}%
  \BibitemOpen
  \bibfield  {author} {\bibinfo {author} {\bibfnamefont {I.}~\bibnamefont
  {Zuriguel}}, \bibinfo {author} {\bibfnamefont {D.~R.}\ \bibnamefont
  {Parisi}}, \bibinfo {author} {\bibfnamefont {R.~C.}\ \bibnamefont {Hidalgo}},
  \bibinfo {author} {\bibfnamefont {C.}~\bibnamefont {Lozano}}, \bibinfo
  {author} {\bibfnamefont {A.}~\bibnamefont {Janda}}, \bibinfo {author}
  {\bibfnamefont {P.~A.}\ \bibnamefont {Gago}}, \bibinfo {author}
  {\bibfnamefont {J.~P.}\ \bibnamefont {Peralta}}, \bibinfo {author}
  {\bibfnamefont {L.~M.}\ \bibnamefont {Ferrer}}, \bibinfo {author}
  {\bibfnamefont {L.~A.}\ \bibnamefont {Pugnaloni}}, \bibinfo {author}
  {\bibfnamefont {E.}~\bibnamefont {Cl{\'e}ment}}, \bibinfo {author}
  {\bibfnamefont {D.}~\bibnamefont {Maza}}, \bibinfo {author} {\bibfnamefont
  {I.}~\bibnamefont {Pagonabarraga}},\ and\ \bibinfo {author} {\bibfnamefont
  {A.}~\bibnamefont {Garcimart{\'i}n}},\ }\bibfield  {title} {\bibinfo {title}
  {Clogging transition of many-particle systems flowing through bottlenecks.},\
  }\href@noop {} {\bibfield  {journal} {\bibinfo  {journal} {Scientific
  reports}\ }\textbf {\bibinfo {volume} {4}},\ \bibinfo {pages} {7324}
  (\bibinfo {year} {2014})}\BibitemShut {NoStop}%
\bibitem [{\citenamefont {To}\ \emph {et~al.}(2001)\citenamefont {To},
  \citenamefont {Lai},\ and\ \citenamefont {Pak}}]{to_jamming_2001}%
  \BibitemOpen
  \bibfield  {author} {\bibinfo {author} {\bibfnamefont {K.}~\bibnamefont
  {To}}, \bibinfo {author} {\bibfnamefont {P.~Y.}\ \bibnamefont {Lai}},\ and\
  \bibinfo {author} {\bibfnamefont {H.~K.}\ \bibnamefont {Pak}},\ }\bibfield
  {title} {\bibinfo {title} {Jamming of granular flow in a two-dimensional
  hopper},\ }\href@noop {} {\bibfield  {journal} {\bibinfo  {journal} {Physical
  Review Letters}\ }\textbf {\bibinfo {volume} {86}},\ \bibinfo {pages} {71}
  (\bibinfo {year} {2001})}\BibitemShut {NoStop}%
\bibitem [{\citenamefont {{Alonso-Marroquin}}\ and\ \citenamefont
  {Mora}(2020)}]{alonso-marroquin_beverloo_2020}%
  \BibitemOpen
  \bibfield  {author} {\bibinfo {author} {\bibfnamefont {F.}~\bibnamefont
  {{Alonso-Marroquin}}}\ and\ \bibinfo {author} {\bibfnamefont
  {P.}~\bibnamefont {Mora}},\ }\bibfield  {title} {\bibinfo {title} {Beverloo
  law for hopper flow derived from self-similar profiles},\ }\href@noop {}
  {\bibfield  {journal} {\bibinfo  {journal} {Granular Matter}\ }\textbf
  {\bibinfo {volume} {23}},\ \bibinfo {pages} {7} (\bibinfo {year}
  {2020})}\BibitemShut {NoStop}%
\bibitem [{\citenamefont {Caitano}\ \emph {et~al.}(2021)\citenamefont
  {Caitano}, \citenamefont {Guerrero}, \citenamefont {Gonzalez}, \citenamefont
  {Zuriguel},\ and\ \citenamefont
  {Garcimartin}}]{caitano_characterization_2021}%
  \BibitemOpen
  \bibfield  {author} {\bibinfo {author} {\bibfnamefont {R.}~\bibnamefont
  {Caitano}}, \bibinfo {author} {\bibfnamefont {B.}~\bibnamefont {Guerrero}},
  \bibinfo {author} {\bibfnamefont {R.}~\bibnamefont {Gonzalez}}, \bibinfo
  {author} {\bibfnamefont {I.}~\bibnamefont {Zuriguel}},\ and\ \bibinfo
  {author} {\bibfnamefont {A.}~\bibnamefont {Garcimartin}},\ }\bibfield
  {title} {\bibinfo {title} {Characterization of the {{Clogging Transition}} in
  {{Vibrated Granular Media}}},\ }\href@noop {} {\bibfield  {journal} {\bibinfo
   {journal} {Physical Review Letters}\ }\textbf {\bibinfo {volume} {127}},\
  \bibinfo {pages} {148002} (\bibinfo {year} {2021})}\BibitemShut {NoStop}%
\bibitem [{\citenamefont {Janda}\ \emph {et~al.}(2008)\citenamefont {Janda},
  \citenamefont {Zuriguel}, \citenamefont {Garcimart{\'i}n}, \citenamefont
  {Pugnaloni},\ and\ \citenamefont {Maza}}]{janda_jamming_2008}%
  \BibitemOpen
  \bibfield  {author} {\bibinfo {author} {\bibfnamefont {A.}~\bibnamefont
  {Janda}}, \bibinfo {author} {\bibfnamefont {I.}~\bibnamefont {Zuriguel}},
  \bibinfo {author} {\bibfnamefont {A.}~\bibnamefont {Garcimart{\'i}n}},
  \bibinfo {author} {\bibfnamefont {L.~A.}\ \bibnamefont {Pugnaloni}},\ and\
  \bibinfo {author} {\bibfnamefont {D.}~\bibnamefont {Maza}},\ }\bibfield
  {title} {\bibinfo {title} {Jamming and critical outlet size in the discharge
  of a two-dimensional silo},\ }\href@noop {} {\bibfield  {journal} {\bibinfo
  {journal} {EPL (Europhysics Letters)}\ }\textbf {\bibinfo {volume} {84}},\
  \bibinfo {pages} {44002} (\bibinfo {year} {2008})}\BibitemShut {NoStop}%
\bibitem [{\citenamefont {Janda}\ \emph {et~al.}(2015)\citenamefont {Janda},
  \citenamefont {Zuriguel}, \citenamefont {Garcimart{\'i}n},\ and\
  \citenamefont {Maza}}]{janda_clogging_2015}%
  \BibitemOpen
  \bibfield  {author} {\bibinfo {author} {\bibfnamefont {A.}~\bibnamefont
  {Janda}}, \bibinfo {author} {\bibfnamefont {I.}~\bibnamefont {Zuriguel}},
  \bibinfo {author} {\bibfnamefont {A.}~\bibnamefont {Garcimart{\'i}n}},\ and\
  \bibinfo {author} {\bibfnamefont {D.}~\bibnamefont {Maza}},\ }\bibfield
  {title} {\bibinfo {title} {Clogging of granular materials in narrow vertical
  pipes discharged at constant velocity},\ }\href@noop {} {\bibfield  {journal}
  {\bibinfo  {journal} {Granular Matter}\ }\textbf {\bibinfo {volume} {17}},\
  \bibinfo {pages} {545} (\bibinfo {year} {2015})}\BibitemShut {NoStop}%
\bibitem [{\citenamefont {Hafez}\ \emph {et~al.}(2021)\citenamefont {Hafez},
  \citenamefont {Liu}, \citenamefont {Finkbeiner}, \citenamefont {Alouhali},
  \citenamefont {Moellendick},\ and\ \citenamefont
  {Santamarina}}]{hafez_effect_2021}%
  \BibitemOpen
  \bibfield  {author} {\bibinfo {author} {\bibfnamefont {A.}~\bibnamefont
  {Hafez}}, \bibinfo {author} {\bibfnamefont {Q.}~\bibnamefont {Liu}}, \bibinfo
  {author} {\bibfnamefont {T.}~\bibnamefont {Finkbeiner}}, \bibinfo {author}
  {\bibfnamefont {R.~A.}\ \bibnamefont {Alouhali}}, \bibinfo {author}
  {\bibfnamefont {T.~E.}\ \bibnamefont {Moellendick}},\ and\ \bibinfo {author}
  {\bibfnamefont {J.~C.}\ \bibnamefont {Santamarina}},\ }\bibfield  {title}
  {\bibinfo {title} {The effect of particle shape on discharge and clogging},\
  }\href@noop {} {\bibfield  {journal} {\bibinfo  {journal} {Scientific
  Reports}\ }\textbf {\bibinfo {volume} {11}},\ \bibinfo {pages} {3309}
  (\bibinfo {year} {2021})}\BibitemShut {NoStop}%
\bibitem [{\citenamefont {Thomas}\ and\ \citenamefont
  {Durian}(2015)}]{thomas_fraction_2015}%
  \BibitemOpen
  \bibfield  {author} {\bibinfo {author} {\bibfnamefont {C.~C.}\ \bibnamefont
  {Thomas}}\ and\ \bibinfo {author} {\bibfnamefont {D.~J.}\ \bibnamefont
  {Durian}},\ }\bibfield  {title} {\bibinfo {title} {Fraction of {{Clogging
  Configurations Sampled}} by {{Granular Hopper Flow}}},\ }\href@noop {}
  {\bibfield  {journal} {\bibinfo  {journal} {Physical Review Letters}\
  }\textbf {\bibinfo {volume} {114}},\ \bibinfo {pages} {178001} (\bibinfo
  {year} {2015})}\BibitemShut {NoStop}%
\bibitem [{\citenamefont {Koivisto}\ and\ \citenamefont
  {Durian}(2017)}]{koivisto_effect_2017}%
  \BibitemOpen
  \bibfield  {author} {\bibinfo {author} {\bibfnamefont {J.}~\bibnamefont
  {Koivisto}}\ and\ \bibinfo {author} {\bibfnamefont {D.~J.}\ \bibnamefont
  {Durian}},\ }\bibfield  {title} {\bibinfo {title} {Effect of interstitial
  fluid on the fraction of flow microstates that precede clogging in granular
  hoppers},\ }\href@noop {} {\bibfield  {journal} {\bibinfo  {journal}
  {Physical Review E}\ }\textbf {\bibinfo {volume} {95}},\ \bibinfo {pages}
  {032904} (\bibinfo {year} {2017})}\BibitemShut {NoStop}%
\bibitem [{\citenamefont {Pong{\'o}}\ \emph {et~al.}(2021)\citenamefont
  {Pong{\'o}}, \citenamefont {Stiga}, \citenamefont {T{\"o}r{\"o}k},
  \citenamefont {L{\'e}vay}, \citenamefont {Szab{\'o}}, \citenamefont
  {Stannarius}, \citenamefont {Hidalgo},\ and\ \citenamefont
  {B{\"o}rzs{\"o}nyi}}]{pongo_flow_2021}%
  \BibitemOpen
  \bibfield  {author} {\bibinfo {author} {\bibfnamefont {T.}~\bibnamefont
  {Pong{\'o}}}, \bibinfo {author} {\bibfnamefont {V.}~\bibnamefont {Stiga}},
  \bibinfo {author} {\bibfnamefont {J.}~\bibnamefont {T{\"o}r{\"o}k}}, \bibinfo
  {author} {\bibfnamefont {S.}~\bibnamefont {L{\'e}vay}}, \bibinfo {author}
  {\bibfnamefont {B.}~\bibnamefont {Szab{\'o}}}, \bibinfo {author}
  {\bibfnamefont {R.}~\bibnamefont {Stannarius}}, \bibinfo {author}
  {\bibfnamefont {R.~C.}\ \bibnamefont {Hidalgo}},\ and\ \bibinfo {author}
  {\bibfnamefont {T.}~\bibnamefont {B{\"o}rzs{\"o}nyi}},\ }\bibfield  {title}
  {\bibinfo {title} {Flow in an hourglass: Particle friction and stiffness
  matter},\ }\href@noop {} {\bibfield  {journal} {\bibinfo  {journal} {New
  Journal of Physics}\ }\textbf {\bibinfo {volume} {23}},\ \bibinfo {pages}
  {023001} (\bibinfo {year} {2021})}\BibitemShut {NoStop}%
\bibitem [{\citenamefont {Hong}\ \emph {et~al.}(2017)\citenamefont {Hong},
  \citenamefont {Kohne}, \citenamefont {Morrell}, \citenamefont {Wang},\ and\
  \citenamefont {Weeks}}]{hong_clogging_2017}%
  \BibitemOpen
  \bibfield  {author} {\bibinfo {author} {\bibfnamefont {X.}~\bibnamefont
  {Hong}}, \bibinfo {author} {\bibfnamefont {M.}~\bibnamefont {Kohne}},
  \bibinfo {author} {\bibfnamefont {M.}~\bibnamefont {Morrell}}, \bibinfo
  {author} {\bibfnamefont {H.}~\bibnamefont {Wang}},\ and\ \bibinfo {author}
  {\bibfnamefont {E.~R.}\ \bibnamefont {Weeks}},\ }\bibfield  {title} {\bibinfo
  {title} {Clogging of soft particles in two-dimensional hoppers},\ }\href@noop
  {} {\bibfield  {journal} {\bibinfo  {journal} {Physical Review E}\ }\textbf
  {\bibinfo {volume} {96}},\ \bibinfo {pages} {062605} (\bibinfo {year}
  {2017})}\BibitemShut {NoStop}%
\bibitem [{\citenamefont {Tao}\ \emph {et~al.}(2021)\citenamefont {Tao},
  \citenamefont {Wilson},\ and\ \citenamefont {Weeks}}]{tao_soft_2021}%
  \BibitemOpen
  \bibfield  {author} {\bibinfo {author} {\bibfnamefont {R.}~\bibnamefont
  {Tao}}, \bibinfo {author} {\bibfnamefont {M.}~\bibnamefont {Wilson}},\ and\
  \bibinfo {author} {\bibfnamefont {E.~R.}\ \bibnamefont {Weeks}},\ }\bibfield
  {title} {\bibinfo {title} {Soft particle clogging in two-dimensional
  hoppers},\ }\href@noop {} {\bibfield  {journal} {\bibinfo  {journal}
  {Physical Review E}\ }\textbf {\bibinfo {volume} {104}},\ \bibinfo {pages}
  {044909} (\bibinfo {year} {2021})}\BibitemShut {NoStop}%
\bibitem [{\citenamefont {Harth}\ \emph {et~al.}(2020)\citenamefont {Harth},
  \citenamefont {Wang}, \citenamefont {Börzsönyi},\ and\ \citenamefont
  {Stannarius}}]{harth2020intermittent}%
  \BibitemOpen
  \bibfield  {author} {\bibinfo {author} {\bibfnamefont {K.}~\bibnamefont
  {Harth}}, \bibinfo {author} {\bibfnamefont {J.}~\bibnamefont {Wang}},
  \bibinfo {author} {\bibfnamefont {T.}~\bibnamefont {Börzsönyi}},\ and\
  \bibinfo {author} {\bibfnamefont {R.}~\bibnamefont {Stannarius}},\ }\bibfield
   {title} {\bibinfo {title} {Intermittent flow and transient congestions of
  soft spheres passing narrow orifices},\ }\href
  {https://doi.org/10.1039/D0SM00938E} {\bibfield  {journal} {\bibinfo
  {journal} {Soft Matter}\ }\textbf {\bibinfo {volume} {16}},\ \bibinfo {pages}
  {8013} (\bibinfo {year} {2020})}\BibitemShut {NoStop}%
\bibitem [{\citenamefont {To}(2005)}]{to_jamming_2005}%
  \BibitemOpen
  \bibfield  {author} {\bibinfo {author} {\bibfnamefont {K.}~\bibnamefont
  {To}},\ }\bibfield  {title} {\bibinfo {title} {Jamming transition in
  two-dimensional hoppers and silos},\ }\href@noop {} {\bibfield  {journal}
  {\bibinfo  {journal} {Phys. Rev. E}\ }\textbf {\bibinfo {volume} {71}},\
  \bibinfo {pages} {060301} (\bibinfo {year} {2005})}\BibitemShut {NoStop}%
\bibitem [{\citenamefont {Zuriguel}\ \emph {et~al.}(2005)\citenamefont
  {Zuriguel}, \citenamefont {Garcimart{\'i}n}, \citenamefont {Maza},
  \citenamefont {Pugnaloni},\ and\ \citenamefont
  {Pastor}}]{zuriguel_jamming_2005}%
  \BibitemOpen
  \bibfield  {author} {\bibinfo {author} {\bibfnamefont {I.}~\bibnamefont
  {Zuriguel}}, \bibinfo {author} {\bibfnamefont {A.}~\bibnamefont
  {Garcimart{\'i}n}}, \bibinfo {author} {\bibfnamefont {D.}~\bibnamefont
  {Maza}}, \bibinfo {author} {\bibfnamefont {L.}~\bibnamefont {Pugnaloni}},\
  and\ \bibinfo {author} {\bibfnamefont {J.}~\bibnamefont {Pastor}},\
  }\bibfield  {title} {\bibinfo {title} {Jamming during the discharge of
  granular matter from a silo},\ }\href@noop {} {\bibfield  {journal} {\bibinfo
   {journal} {Physical Review E}\ }\textbf {\bibinfo {volume} {71}},\ \bibinfo
  {pages} {051303} (\bibinfo {year} {2005})}\BibitemShut {NoStop}%
\bibitem [{\citenamefont {Tang}\ \emph {et~al.}(2009)\citenamefont {Tang},
  \citenamefont {Sagdiphour},\ and\ \citenamefont {Behringer}}]{Tang09}%
  \BibitemOpen
  \bibfield  {author} {\bibinfo {author} {\bibfnamefont {J.}~\bibnamefont
  {Tang}}, \bibinfo {author} {\bibfnamefont {S.}~\bibnamefont {Sagdiphour}},\
  and\ \bibinfo {author} {\bibfnamefont {R.~P.}\ \bibnamefont {Behringer}},\
  }\bibfield  {title} {\bibinfo {title} {{Jamming and Flow in 2D Hoppers}},\
  }\href@noop {} {\bibfield  {journal} {\bibinfo  {journal} {AIP Conf. Proc.}\
  }\textbf {\bibinfo {volume} {1145}},\ \bibinfo {pages} {515} (\bibinfo {year}
  {2009})}\BibitemShut {NoStop}%
\bibitem [{\citenamefont {Cubuk}\ \emph {et~al.}(2015)\citenamefont {Cubuk},
  \citenamefont {Schoenholz}, \citenamefont {Rieser}, \citenamefont {Malone},
  \citenamefont {Rottler}, \citenamefont {Durian}, \citenamefont {Kaxiras},\
  and\ \citenamefont {Liu}}]{cubuk_identifying_2015}%
  \BibitemOpen
  \bibfield  {author} {\bibinfo {author} {\bibfnamefont {E.}~\bibnamefont
  {Cubuk}}, \bibinfo {author} {\bibfnamefont {S.}~\bibnamefont {Schoenholz}},
  \bibinfo {author} {\bibfnamefont {J.}~\bibnamefont {Rieser}}, \bibinfo
  {author} {\bibfnamefont {B.}~\bibnamefont {Malone}}, \bibinfo {author}
  {\bibfnamefont {J.}~\bibnamefont {Rottler}}, \bibinfo {author} {\bibfnamefont
  {D.}~\bibnamefont {Durian}}, \bibinfo {author} {\bibfnamefont
  {E.}~\bibnamefont {Kaxiras}},\ and\ \bibinfo {author} {\bibfnamefont
  {A.}~\bibnamefont {Liu}},\ }\bibfield  {title} {\bibinfo {title} {Identifying
  {{Structural Flow Defects}} in {{Disordered Solids Using Machine-Learning
  Methods}}},\ }\href@noop {} {\bibfield  {journal} {\bibinfo  {journal}
  {Physical Review Letters}\ }\textbf {\bibinfo {volume} {114}},\ \bibinfo
  {pages} {108001} (\bibinfo {year} {2015})}\BibitemShut {NoStop}%
\bibitem [{\citenamefont {Cubuk}\ \emph {et~al.}(2017)\citenamefont {Cubuk},
  \citenamefont {Ivancic}, \citenamefont {Schoenholz}, \citenamefont
  {Strickland}, \citenamefont {Basu}, \citenamefont {Davidson}, \citenamefont
  {Fontaine}, \citenamefont {Hor}, \citenamefont {Huang}, \citenamefont
  {Jiang}, \citenamefont {Keim}, \citenamefont {Koshigan}, \citenamefont
  {Lefever}, \citenamefont {Liu}, \citenamefont {Ma}, \citenamefont
  {Magagnosc}, \citenamefont {Morrow}, \citenamefont {Ortiz}, \citenamefont
  {Rieser}, \citenamefont {Shavit}, \citenamefont {Still}, \citenamefont {Xu},
  \citenamefont {Zhang}, \citenamefont {Nordstrom}, \citenamefont {Arratia},
  \citenamefont {Carpick}, \citenamefont {Durian}, \citenamefont {Fakhraai},
  \citenamefont {Jerolmack}, \citenamefont {Lee}, \citenamefont {Li},
  \citenamefont {Riggleman}, \citenamefont {Turner}, \citenamefont {Yodh},
  \citenamefont {Gianola},\ and\ \citenamefont
  {Liu}}]{cubuk_structureproperty_2017}%
  \BibitemOpen
  \bibfield  {author} {\bibinfo {author} {\bibfnamefont {E.~D.}\ \bibnamefont
  {Cubuk}}, \bibinfo {author} {\bibfnamefont {R.~J.~S.}\ \bibnamefont
  {Ivancic}}, \bibinfo {author} {\bibfnamefont {S.~S.}\ \bibnamefont
  {Schoenholz}}, \bibinfo {author} {\bibfnamefont {D.~J.}\ \bibnamefont
  {Strickland}}, \bibinfo {author} {\bibfnamefont {A.}~\bibnamefont {Basu}},
  \bibinfo {author} {\bibfnamefont {Z.~S.}\ \bibnamefont {Davidson}}, \bibinfo
  {author} {\bibfnamefont {J.}~\bibnamefont {Fontaine}}, \bibinfo {author}
  {\bibfnamefont {J.~L.}\ \bibnamefont {Hor}}, \bibinfo {author} {\bibfnamefont
  {Y.-R.}\ \bibnamefont {Huang}}, \bibinfo {author} {\bibfnamefont
  {Y.}~\bibnamefont {Jiang}}, \bibinfo {author} {\bibfnamefont {N.~C.}\
  \bibnamefont {Keim}}, \bibinfo {author} {\bibfnamefont {K.~D.}\ \bibnamefont
  {Koshigan}}, \bibinfo {author} {\bibfnamefont {J.~A.}\ \bibnamefont
  {Lefever}}, \bibinfo {author} {\bibfnamefont {T.}~\bibnamefont {Liu}},
  \bibinfo {author} {\bibfnamefont {X.-G.}\ \bibnamefont {Ma}}, \bibinfo
  {author} {\bibfnamefont {D.~J.}\ \bibnamefont {Magagnosc}}, \bibinfo {author}
  {\bibfnamefont {E.}~\bibnamefont {Morrow}}, \bibinfo {author} {\bibfnamefont
  {C.~P.}\ \bibnamefont {Ortiz}}, \bibinfo {author} {\bibfnamefont {J.~M.}\
  \bibnamefont {Rieser}}, \bibinfo {author} {\bibfnamefont {A.}~\bibnamefont
  {Shavit}}, \bibinfo {author} {\bibfnamefont {T.}~\bibnamefont {Still}},
  \bibinfo {author} {\bibfnamefont {Y.}~\bibnamefont {Xu}}, \bibinfo {author}
  {\bibfnamefont {Y.}~\bibnamefont {Zhang}}, \bibinfo {author} {\bibfnamefont
  {K.~N.}\ \bibnamefont {Nordstrom}}, \bibinfo {author} {\bibfnamefont {P.~E.}\
  \bibnamefont {Arratia}}, \bibinfo {author} {\bibfnamefont {R.~W.}\
  \bibnamefont {Carpick}}, \bibinfo {author} {\bibfnamefont {D.~J.}\
  \bibnamefont {Durian}}, \bibinfo {author} {\bibfnamefont {Z.}~\bibnamefont
  {Fakhraai}}, \bibinfo {author} {\bibfnamefont {D.~J.}\ \bibnamefont
  {Jerolmack}}, \bibinfo {author} {\bibfnamefont {D.}~\bibnamefont {Lee}},
  \bibinfo {author} {\bibfnamefont {J.}~\bibnamefont {Li}}, \bibinfo {author}
  {\bibfnamefont {R.}~\bibnamefont {Riggleman}}, \bibinfo {author}
  {\bibfnamefont {K.~T.}\ \bibnamefont {Turner}}, \bibinfo {author}
  {\bibfnamefont {A.~G.}\ \bibnamefont {Yodh}}, \bibinfo {author}
  {\bibfnamefont {D.~S.}\ \bibnamefont {Gianola}},\ and\ \bibinfo {author}
  {\bibfnamefont {A.~J.}\ \bibnamefont {Liu}},\ }\bibfield  {title} {\bibinfo
  {title} {Structure-property relationships from universal signatures of
  plasticity in disordered solids},\ }\href
  {https://doi.org/10.1126/science.aai8830} {\bibfield  {journal} {\bibinfo
  {journal} {Science}\ }\textbf {\bibinfo {volume} {358}},\ \bibinfo {pages}
  {1033} (\bibinfo {year} {2017})},\ \bibinfo {note} {publisher: American
  Association for the Advancement of Science}\BibitemShut {NoStop}%
\bibitem [{\citenamefont {Xiao}\ \emph {et~al.}(2023)\citenamefont {Xiao},
  \citenamefont {Zhang}, \citenamefont {Yang}, \citenamefont {Ivancic},
  \citenamefont {Ridout}, \citenamefont {Riggleman}, \citenamefont {Durian},\
  and\ \citenamefont {Liu}}]{xiao_identifying_2023}%
  \BibitemOpen
  \bibfield  {author} {\bibinfo {author} {\bibfnamefont {H.}~\bibnamefont
  {Xiao}}, \bibinfo {author} {\bibfnamefont {G.}~\bibnamefont {Zhang}},
  \bibinfo {author} {\bibfnamefont {E.}~\bibnamefont {Yang}}, \bibinfo {author}
  {\bibfnamefont {R.}~\bibnamefont {Ivancic}}, \bibinfo {author} {\bibfnamefont
  {S.}~\bibnamefont {Ridout}}, \bibinfo {author} {\bibfnamefont
  {R.}~\bibnamefont {Riggleman}}, \bibinfo {author} {\bibfnamefont {D.~J.}\
  \bibnamefont {Durian}},\ and\ \bibinfo {author} {\bibfnamefont {A.~J.}\
  \bibnamefont {Liu}},\ }\bibfield  {title} {\bibinfo {title} {Identifying
  microscopic factors that influence ductility in disordered solids},\
  }\href@noop {} {\bibfield  {journal} {\bibinfo  {journal} {Proceedings of the
  National Academy of Sciences}\ }\textbf {\bibinfo {volume} {120}},\ \bibinfo
  {pages} {e2307552120} (\bibinfo {year} {2023})}\BibitemShut {NoStop}%
\bibitem [{\citenamefont {Hathcock}\ \emph {et~al.}(2023)\citenamefont
  {Hathcock}, \citenamefont {Dillavou}, \citenamefont {Hanlan}, \citenamefont
  {Durian},\ and\ \citenamefont {Tu}}]{hathcock_stochastic_2023}%
  \BibitemOpen
  \bibfield  {author} {\bibinfo {author} {\bibfnamefont {D.}~\bibnamefont
  {Hathcock}}, \bibinfo {author} {\bibfnamefont {S.}~\bibnamefont {Dillavou}},
  \bibinfo {author} {\bibfnamefont {J.~M.}\ \bibnamefont {Hanlan}}, \bibinfo
  {author} {\bibfnamefont {D.~J.}\ \bibnamefont {Durian}},\ and\ \bibinfo
  {author} {\bibfnamefont {Y.}~\bibnamefont {Tu}},\ }\href@noop {} {\bibinfo
  {title} {Stochastic dynamics of granular hopper flows: A slow hidden mode
  controls the stability of clogs}} (\bibinfo {year} {2023}),\ \Eprint
  {https://arxiv.org/abs/2312.01194} {{arXiv}:2312.01194} \BibitemShut
  {NoStop}%
\bibitem [{Dat()}]{DataInDryadRepository}%
  \BibitemOpen
  \href@noop {} {}\bibinfo {note}
  {\url{https://doi.org/10.5061/dryad.cvdncjtb5} (link not yet
  live)}\BibitemShut {NoStop}%
\bibitem [{Pyt()}]{PythonScript}%
  \BibitemOpen
  \href@noop {} {}\bibinfo {note}
  {\url{https://doi.org/10.5281/zenodo.10895419} (link not yet
  live)}\BibitemShut {NoStop}%
\bibitem [{\citenamefont {Burges}(1998)}]{burges_tutorial_1998}%
  \BibitemOpen
  \bibfield  {author} {\bibinfo {author} {\bibfnamefont {C.~J.}\ \bibnamefont
  {Burges}},\ }\bibfield  {title} {\bibinfo {title} {A {{Tutorial}} on
  {{Support Vector Machines}} for {{Pattern Recognition}}},\ }\href@noop {}
  {\bibfield  {journal} {\bibinfo  {journal} {Data Mining and Knowledge
  Discovery}\ }\textbf {\bibinfo {volume} {2}},\ \bibinfo {pages} {121}
  (\bibinfo {year} {1998})}\BibitemShut {NoStop}%
\bibitem [{\citenamefont {LeCun}\ \emph {et~al.}(2015)\citenamefont {LeCun},
  \citenamefont {Bengio},\ and\ \citenamefont {Hinton}}]{lecun_deep_2015}%
  \BibitemOpen
  \bibfield  {author} {\bibinfo {author} {\bibfnamefont {Y.}~\bibnamefont
  {LeCun}}, \bibinfo {author} {\bibfnamefont {Y.}~\bibnamefont {Bengio}},\ and\
  \bibinfo {author} {\bibfnamefont {G.}~\bibnamefont {Hinton}},\ }\bibfield
  {title} {\bibinfo {title} {Deep learning},\ }\href@noop {} {\bibfield
  {journal} {\bibinfo  {journal} {Nature}\ }\textbf {\bibinfo {volume} {521}},\
  \bibinfo {pages} {436} (\bibinfo {year} {2015})}\BibitemShut {NoStop}%
\bibitem [{\citenamefont {Li}\ \emph {et~al.}(2022)\citenamefont {Li},
  \citenamefont {Liu}, \citenamefont {Yang}, \citenamefont {Peng},\ and\
  \citenamefont {Zhou}}]{li_survey_2022}%
  \BibitemOpen
  \bibfield  {author} {\bibinfo {author} {\bibfnamefont {Z.}~\bibnamefont
  {Li}}, \bibinfo {author} {\bibfnamefont {F.}~\bibnamefont {Liu}}, \bibinfo
  {author} {\bibfnamefont {W.}~\bibnamefont {Yang}}, \bibinfo {author}
  {\bibfnamefont {S.}~\bibnamefont {Peng}},\ and\ \bibinfo {author}
  {\bibfnamefont {J.}~\bibnamefont {Zhou}},\ }\bibfield  {title} {\bibinfo
  {title} {A {{Survey}} of {{Convolutional Neural Networks}}: {{Analysis}},
  {{Applications}}, and {{Prospects}}},\ }\href@noop {} {\bibfield  {journal}
  {\bibinfo  {journal} {IEEE Transactions on Neural Networks and Learning
  Systems}\ }\textbf {\bibinfo {volume} {33}},\ \bibinfo {pages} {6999}
  (\bibinfo {year} {2022})}\BibitemShut {NoStop}%
\bibitem [{\citenamefont {Rosasco}\ \emph {et~al.}(2004)\citenamefont
  {Rosasco}, \citenamefont {Vito}, \citenamefont {Caponnetto}, \citenamefont
  {Piana},\ and\ \citenamefont {Verri}}]{rosasco_are_2004}%
  \BibitemOpen
  \bibfield  {author} {\bibinfo {author} {\bibfnamefont {L.}~\bibnamefont
  {Rosasco}}, \bibinfo {author} {\bibfnamefont {E.~D.}\ \bibnamefont {Vito}},
  \bibinfo {author} {\bibfnamefont {A.}~\bibnamefont {Caponnetto}}, \bibinfo
  {author} {\bibfnamefont {M.}~\bibnamefont {Piana}},\ and\ \bibinfo {author}
  {\bibfnamefont {A.}~\bibnamefont {Verri}},\ }\bibfield  {title} {\bibinfo
  {title} {Are {{Loss Functions All}} the {{Same}}?},\ }\href@noop {}
  {\bibfield  {journal} {\bibinfo  {journal} {Neural Computation}\ }\textbf
  {\bibinfo {volume} {16}},\ \bibinfo {pages} {1063} (\bibinfo {year}
  {2004})}\BibitemShut {NoStop}%
\bibitem [{\citenamefont {{de Boer}}\ \emph {et~al.}(2005)\citenamefont {{de
  Boer}}, \citenamefont {Kroese}, \citenamefont {Mannor},\ and\ \citenamefont
  {Rubinstein}}]{de_boer_tutorial_2005}%
  \BibitemOpen
  \bibfield  {author} {\bibinfo {author} {\bibfnamefont {P.-T.}\ \bibnamefont
  {{de Boer}}}, \bibinfo {author} {\bibfnamefont {D.~P.}\ \bibnamefont
  {Kroese}}, \bibinfo {author} {\bibfnamefont {S.}~\bibnamefont {Mannor}},\
  and\ \bibinfo {author} {\bibfnamefont {R.~Y.}\ \bibnamefont {Rubinstein}},\
  }\bibfield  {title} {\bibinfo {title} {A {{Tutorial}} on the {{Cross-Entropy
  Method}}},\ }\href@noop {} {\bibfield  {journal} {\bibinfo  {journal} {Annals
  of Operations Research}\ }\textbf {\bibinfo {volume} {134}},\ \bibinfo
  {pages} {19} (\bibinfo {year} {2005})}\BibitemShut {NoStop}%
\bibitem [{\citenamefont {Behler}\ and\ \citenamefont
  {Parrinello}(2007)}]{behler_generalized_2007}%
  \BibitemOpen
  \bibfield  {author} {\bibinfo {author} {\bibfnamefont {J.}~\bibnamefont
  {Behler}}\ and\ \bibinfo {author} {\bibfnamefont {M.}~\bibnamefont
  {Parrinello}},\ }\bibfield  {title} {\bibinfo {title} {Generalized
  {{Neural-Network Representation}} of {{High-Dimensional Potential-Energy
  Surfaces}}},\ }\href@noop {} {\bibfield  {journal} {\bibinfo  {journal}
  {Physical Review Letters}\ }\textbf {\bibinfo {volume} {98}},\ \bibinfo
  {pages} {146401} (\bibinfo {year} {2007})}\BibitemShut {NoStop}%
\bibitem [{\citenamefont {Schaeffer}\ \emph {et~al.}(2023)\citenamefont
  {Schaeffer}, \citenamefont {Khona}, \citenamefont {Robertson}, \citenamefont
  {Boopathy}, \citenamefont {Pistunova}, \citenamefont {Rocks}, \citenamefont
  {Fiete},\ and\ \citenamefont {Koyejo}}]{schaeffer_double_2023}%
  \BibitemOpen
  \bibfield  {author} {\bibinfo {author} {\bibfnamefont {R.}~\bibnamefont
  {Schaeffer}}, \bibinfo {author} {\bibfnamefont {M.}~\bibnamefont {Khona}},
  \bibinfo {author} {\bibfnamefont {Z.}~\bibnamefont {Robertson}}, \bibinfo
  {author} {\bibfnamefont {A.}~\bibnamefont {Boopathy}}, \bibinfo {author}
  {\bibfnamefont {K.}~\bibnamefont {Pistunova}}, \bibinfo {author}
  {\bibfnamefont {J.~W.}\ \bibnamefont {Rocks}}, \bibinfo {author}
  {\bibfnamefont {I.~R.}\ \bibnamefont {Fiete}},\ and\ \bibinfo {author}
  {\bibfnamefont {O.}~\bibnamefont {Koyejo}},\ }\href@noop {} {\bibinfo {title}
  {Double {{Descent Demystified}}: {{Identifying}}, {{Interpreting}} \&
  {{Ablating}} the {{Sources}} of a {{Deep Learning Puzzle}}}} (\bibinfo {year}
  {2023}),\ \Eprint {https://arxiv.org/abs/2303.14151} {{arXiv}:2303.14151}
  \BibitemShut {NoStop}%
\bibitem [{\citenamefont {Dillavou}\ \emph {et~al.}(2022)\citenamefont
  {Dillavou}, \citenamefont {Bar-Sinai}, \citenamefont {Brenner},\ and\
  \citenamefont {Rubinstein}}]{dillavou_quality_2022}%
  \BibitemOpen
  \bibfield  {author} {\bibinfo {author} {\bibfnamefont {S.}~\bibnamefont
  {Dillavou}}, \bibinfo {author} {\bibfnamefont {Y.}~\bibnamefont {Bar-Sinai}},
  \bibinfo {author} {\bibfnamefont {M.~P.}\ \bibnamefont {Brenner}},\ and\
  \bibinfo {author} {\bibfnamefont {S.~M.}\ \bibnamefont {Rubinstein}},\
  }\bibfield  {title} {\bibinfo {title} {Beyond quality and quantity:
  {{Spatial}} distribution of contact encodes frictional strength},\ }\href
  {https://doi.org/10.1103/PhysRevE.106.L033001} {\bibfield  {journal}
  {\bibinfo  {journal} {Physical Review E}\ }\textbf {\bibinfo {volume}
  {106}},\ \bibinfo {pages} {L033001} (\bibinfo {year} {2022})}\BibitemShut
  {NoStop}%
\bibitem [{\citenamefont {G{\'e}ron}(2019)}]{geron_hands-machine_2019}%
  \BibitemOpen
  \bibfield  {author} {\bibinfo {author} {\bibfnamefont {A.}~\bibnamefont
  {G{\'e}ron}},\ }\href@noop {} {\emph {\bibinfo {title} {Hands-on Machine
  Learning with {{Scikit-Learn}}, {{Keras}}, and {{TensorFlow}}: Concepts,
  Tools, and Techniques to Build Intelligent Systems}}},\ \bibinfo {edition}
  {second edition}\ ed.\ (\bibinfo  {publisher} {O'Reilly Media, Inc},\
  \bibinfo {address} {Beijing [China] ; Sebastopol, CA},\ \bibinfo {year}
  {2019})\BibitemShut {NoStop}%
\bibitem [{\citenamefont {Bishop}(2006)}]{bishop_pattern_2006}%
  \BibitemOpen
  \bibfield  {author} {\bibinfo {author} {\bibfnamefont {C.~M.}\ \bibnamefont
  {Bishop}},\ }\href@noop {} {\emph {\bibinfo {title} {Pattern Recognition and
  Machine Learning}}},\ Information Science and Statistics\ (\bibinfo
  {publisher} {Springer},\ \bibinfo {address} {New York},\ \bibinfo {year}
  {2006})\BibitemShut {NoStop}%
\bibitem [{\citenamefont {Harrington}\ \emph {et~al.}(2019)\citenamefont
  {Harrington}, \citenamefont {Liu},\ and\ \citenamefont
  {Durian}}]{harrington_machine_2019}%
  \BibitemOpen
  \bibfield  {author} {\bibinfo {author} {\bibfnamefont {M.}~\bibnamefont
  {Harrington}}, \bibinfo {author} {\bibfnamefont {A.~J.}\ \bibnamefont
  {Liu}},\ and\ \bibinfo {author} {\bibfnamefont {D.~J.}\ \bibnamefont
  {Durian}},\ }\bibfield  {title} {\bibinfo {title} {Machine learning
  characterization of structural defects in amorphous packings of dimers and
  ellipses},\ }\href@noop {} {\bibfield  {journal} {\bibinfo  {journal}
  {Physical Review E}\ }\textbf {\bibinfo {volume} {99}},\ \bibinfo {pages}
  {022903} (\bibinfo {year} {2019})}\BibitemShut {NoStop}%
\bibitem [{\citenamefont {Schoenholz}\ \emph {et~al.}(2016)\citenamefont
  {Schoenholz}, \citenamefont {Cubuk}, \citenamefont {Sussman}, \citenamefont
  {Kaxiras},\ and\ \citenamefont {Liu}}]{schoenholz_structural_2016}%
  \BibitemOpen
  \bibfield  {author} {\bibinfo {author} {\bibfnamefont {S.}~\bibnamefont
  {Schoenholz}}, \bibinfo {author} {\bibfnamefont {E.}~\bibnamefont {Cubuk}},
  \bibinfo {author} {\bibfnamefont {D.}~\bibnamefont {Sussman}}, \bibinfo
  {author} {\bibfnamefont {E.}~\bibnamefont {Kaxiras}},\ and\ \bibinfo {author}
  {\bibfnamefont {A.}~\bibnamefont {Liu}},\ }\bibfield  {title} {\bibinfo
  {title} {A structural approach to relaxation in glassy liquids},\ }\href@noop
  {} {\bibfield  {journal} {\bibinfo  {journal} {Nature Physics}\ }\textbf
  {\bibinfo {volume} {12}},\ \bibinfo {pages} {469} (\bibinfo {year}
  {2016})}\BibitemShut {NoStop}%
\bibitem [{\citenamefont {Rocks}\ \emph {et~al.}(2021)\citenamefont {Rocks},
  \citenamefont {Ridout},\ and\ \citenamefont
  {Liu}}]{rocks_learning-based_2021}%
  \BibitemOpen
  \bibfield  {author} {\bibinfo {author} {\bibfnamefont {J.~W.}\ \bibnamefont
  {Rocks}}, \bibinfo {author} {\bibfnamefont {S.~A.}\ \bibnamefont {Ridout}},\
  and\ \bibinfo {author} {\bibfnamefont {A.~J.}\ \bibnamefont {Liu}},\
  }\bibfield  {title} {\bibinfo {title} {Learning-based approach to plasticity
  in athermal sheared amorphous packings: {{Improving}} softness},\ }\href@noop
  {} {\bibfield  {journal} {\bibinfo  {journal} {APL Materials}\ }\textbf
  {\bibinfo {volume} {9}} (\bibinfo {year} {2021})}\BibitemShut {NoStop}%
\bibitem [{\citenamefont {Battaglia}\ \emph {et~al.}(2018)\citenamefont
  {Battaglia}, \citenamefont {Hamrick}, \citenamefont {Bapst}, \citenamefont
  {{Sanchez-Gonzalez}}, \citenamefont {Zambaldi}, \citenamefont {Malinowski},
  \citenamefont {Tacchetti}, \citenamefont {Raposo}, \citenamefont {Santoro},
  \citenamefont {Faulkner}, \citenamefont {Gulcehre}, \citenamefont {Song},
  \citenamefont {Ballard}, \citenamefont {Gilmer}, \citenamefont {Dahl},
  \citenamefont {Vaswani}, \citenamefont {Allen}, \citenamefont {Nash},
  \citenamefont {Langston}, \citenamefont {Dyer}, \citenamefont {Heess},
  \citenamefont {Wierstra}, \citenamefont {Kohli}, \citenamefont {Botvinick},
  \citenamefont {Vinyals}, \citenamefont {Li},\ and\ \citenamefont
  {Pascanu}}]{battaglia_relational_2018}%
  \BibitemOpen
  \bibfield  {author} {\bibinfo {author} {\bibfnamefont {P.~W.}\ \bibnamefont
  {Battaglia}}, \bibinfo {author} {\bibfnamefont {J.~B.}\ \bibnamefont
  {Hamrick}}, \bibinfo {author} {\bibfnamefont {V.}~\bibnamefont {Bapst}},
  \bibinfo {author} {\bibfnamefont {A.}~\bibnamefont {{Sanchez-Gonzalez}}},
  \bibinfo {author} {\bibfnamefont {V.}~\bibnamefont {Zambaldi}}, \bibinfo
  {author} {\bibfnamefont {M.}~\bibnamefont {Malinowski}}, \bibinfo {author}
  {\bibfnamefont {A.}~\bibnamefont {Tacchetti}}, \bibinfo {author}
  {\bibfnamefont {D.}~\bibnamefont {Raposo}}, \bibinfo {author} {\bibfnamefont
  {A.}~\bibnamefont {Santoro}}, \bibinfo {author} {\bibfnamefont
  {R.}~\bibnamefont {Faulkner}}, \bibinfo {author} {\bibfnamefont
  {C.}~\bibnamefont {Gulcehre}}, \bibinfo {author} {\bibfnamefont
  {F.}~\bibnamefont {Song}}, \bibinfo {author} {\bibfnamefont {A.}~\bibnamefont
  {Ballard}}, \bibinfo {author} {\bibfnamefont {J.}~\bibnamefont {Gilmer}},
  \bibinfo {author} {\bibfnamefont {G.}~\bibnamefont {Dahl}}, \bibinfo {author}
  {\bibfnamefont {A.}~\bibnamefont {Vaswani}}, \bibinfo {author} {\bibfnamefont
  {K.}~\bibnamefont {Allen}}, \bibinfo {author} {\bibfnamefont
  {C.}~\bibnamefont {Nash}}, \bibinfo {author} {\bibfnamefont {V.}~\bibnamefont
  {Langston}}, \bibinfo {author} {\bibfnamefont {C.}~\bibnamefont {Dyer}},
  \bibinfo {author} {\bibfnamefont {N.}~\bibnamefont {Heess}}, \bibinfo
  {author} {\bibfnamefont {D.}~\bibnamefont {Wierstra}}, \bibinfo {author}
  {\bibfnamefont {P.}~\bibnamefont {Kohli}}, \bibinfo {author} {\bibfnamefont
  {M.}~\bibnamefont {Botvinick}}, \bibinfo {author} {\bibfnamefont
  {O.}~\bibnamefont {Vinyals}}, \bibinfo {author} {\bibfnamefont
  {Y.}~\bibnamefont {Li}},\ and\ \bibinfo {author} {\bibfnamefont
  {R.}~\bibnamefont {Pascanu}},\ }\href@noop {} {\bibinfo {title} {Relational
  inductive biases, deep learning, and graph networks}} (\bibinfo {year}
  {2018}),\ \Eprint {https://arxiv.org/abs/1806.01261} {{arXiv}:1806.01261}
  \BibitemShut {NoStop}%
\end{thebibliography}%

\end{document}